 \documentclass[final,3p,times]{elsarticle}

 \usepackage{graphics}
 \usepackage{graphicx}
\usepackage{epstopdf}
\usepackage{color}

\usepackage{amssymb}
\usepackage{multicol}

\usepackage{subfigure} 

\usepackage{preprintcover}
\PreprintIdNumber{}
\PreprintCoverPaperTitle{Transverse momentum spectra of charged particles in proton-proton collisions  at $\sqrt{s} = 900$~GeV with ALICE at the LHC}
\PreprintCoverAbstract{The inclusive charged particle 
transverse momentum distribution 
is measured in proton-proton collisions at $\sqrt{s} = 900$~GeV at 
the LHC using 
the ALICE detector. 
The measurement is performed in the central pseudorapidity 
region $(|\eta|<0.8)$
over the transverse momentum range $0.15<p_T<10$~GeV/$c$. The correlation 
between transverse momentum and particle multiplicity is also studied. 
Results are presented for inelastic (INEL) and non-single-diffractive (NSD)
events. 
The average transverse momentum for $|\eta|<0.8$ is 
\mpt$_{\rm INEL}=0.483\pm0.001$~(stat.)~$\pm0.007$~(syst.)~GeV/$c$ and  
\mpt$_{\rm NSD}=0.489\pm0.001$~(stat.)~$\pm0.007$~(syst.)~GeV/$c$,
respectively.
The data exhibit a slightly
larger \mpt 
than measurements in 
wider pseudorapidity
intervals. The results are compared to 
simulations with the Monte Carlo event generators PYTHIA and PHOJET. }
\PreprintJournalName{PLB}

\usepackage{lineno}
\usepackage[english]{babel}

 \makeatletter

\newenvironment{figurehere}
{\def\@captype{figure}}
{}
\makeatother

\def\mpt{$\left<p_T\right>$~}
\def\mbor{MB$_{\rm OR}$~}
\def\mband{MB$_{\rm AND}$~}

\journal{Physics Letters B}

\begin{document}

\begin{frontmatter}



\title{Transverse momentum spectra of charged particles in proton-proton collisions  at $\sqrt{s} = 900$~GeV with ALICE at the LHC\\~\\
ALICE Collaboration}




\author[77]{K.~Aamodt}
\author[43]{N.~Abel}
\author[75]{U.~Abeysekara}
\author[42]{A.~Abrahantes~Quintana}
\author[112]{A.~Abramyan}
\author[85]{D.~Adamov\'{a}}
\author[25]{M.M.~Aggarwal}
\author[40]{G.~Aglieri~Rinella}
\author[18]{A.G.~Agocs}
\author[63]{S.~Aguilar~Salazar}
% ...
\fntext[f1]{Also at{ Laboratoire de Physique Corpusculaire (LPC), Clermont Universit\'{e}, Universit\'{e} Blaise Pascal, CNRS--IN2P3, Clermont-Ferrand, France}}
\fntext[f2]{Also at{ Frankfurt Institute for Advanced Studies, Johann Wolfgang Goethe-Universit\"{a}t Frankfurt, Frankfurt, Germany}}
\fntext[f3]{Now at{ Sezione INFN, Padova, Italy}}
\fntext[f4]{Now at{ Research Division and ExtreMe Matter Institute EMMI, GSI Helmholtzzentrum f\"{u}r Schwerionenforschung, Darmstadt, Germany}}
\fntext[f5]{Now at{ Institut f\"{u}r Kernphysik, Johann Wolfgang Goethe-Universit\"{a}t Frankfurt, Frankfurt, Germany}}
\fntext[f6]{Now at{ Physics Department, University of Cape Town, iThemba Laboratories, Cape Town, South Africa}}
\fntext[f7]{Now at{ National Institute for Physics and Nuclear Engineering, Bucharest, Romania}}
\fntext[f8]{Also at{ University of Houston, Houston, TX, United States}}
\fntext[f9]{Now at{ Faculty of Science, P.J.~\v{S}af\'{a}rik University, Ko\v{s}ice, Slovakia}}
\fntext[f10]{Now at{ European Organization for Nuclear Research (CERN), Geneva, Switzerland}}
\fntext[f11]{Now at{ Helsinki Institute of Physics (HIP) and University of Jyv\"{a}skyl\"{a}, Jyv\"{a}skyl\"{a}, Finland}}
\fntext[f12]{Now at{ Institut Pluridisciplinaire Hubert Curien (IPHC), Universit\'{e} de Strasbourg, CNRS-IN2P3, Strasbourg, France}}
\fntext[f13]{Now at{ Sezione INFN, Bari, Italy}}
\fntext[f14]{Now at{ Institut f\"{u}r Kernphysik, Westf\"{a}lische Wilhelms-Universit\"{a}t M\"{u}nster, M\"{u}nster, Germany}}
\fntext[f15]{Now at: University of Technology and Austrian Academy of Sciences, Vienna, Austria}
\fntext[f16]{Also at{ Lawrence Livermore National Laboratory, Livermore, CA, United States}}
\fntext[f17]{Also at{ European Organization for Nuclear Research (CERN), Geneva, Switzerland}}
\fntext[f18]{Now at { Secci\'{o}n F\'{\i}sica, Departamento de Ciencias, Pontificia Universidad Cat\'{o}lica del Per\'{u}, Lima, Peru}}
\fntext[f19]{Deceased}
\fntext[f20]{Now at{ Yale University, New Haven, CT, United States}}
\fntext[f21]{Now at{ University of Tsukuba, Tsukuba, Japan}}
\fntext[f22]{Also at { Centro Fermi -- Centro Studi e Ricerche e Museo Storico della Fisica ``Enrico Fermi'', Rome, Italy}}
\fntext[f23]{Now at{ Dipartimento Interateneo di Fisica `M.~Merlin' and Sezione INFN, Bari, Italy}}
\fntext[f24]{Also at{ Laboratoire de Physique Subatomique et de Cosmologie (LPSC), Universit\'{e} Joseph Fourier, CNRS-IN2P3, Institut Polytechnique de Grenoble, Grenoble, France}}
\fntext[f25]{Now at{ Dipartimento di Fisica Sperimentale dell'Universit\`{a} and Sezione INFN, Turin, Italy}}
\fntext[f26]{Now at{ Physics Department, Creighton University, Omaha, NE, United States}}
\fntext[f27]{Now at{ Commissariat \`{a} l'Energie Atomique, IRFU, Saclay, France}}
\fntext[f28]{Also at{ Department of Physics, University of Oslo, Oslo, Norway}}
\fntext[f29]{Now at{ Physikalisches Institut, Ruprecht-Karls-Universit\"{a}t Heidelberg, Heidelberg, Germany}}
\fntext[f30]{Now at{ Institut f\"{u}r Kernphysik, Technische Universit\"{a}t Darmstadt, Darmstadt, Germany}}
\fntext[f31]{Now at{ Department of Physics and Technology, University of Bergen, Bergen, Norway}}
\fntext[f32]{Now at{ Physics Department, University of Athens, Athens, Greece}}
\fntext[f33]{Also at{ Institut f\"{u}r Kernphysik, Westf\"{a}lische Wilhelms-Universit\"{a}t M\"{u}nster, M\"{u}nster, Germany}}
\fntext[f34]{Now at{ SUBATECH, Ecole des Mines de Nantes, Universit\'{e} de Nantes, CNRS-IN2P3, Nantes, France}}
\fntext[f35]{Now at{ Universit\'{e} de Lyon, Universit\'{e} Lyon 1, CNRS/IN2P3, IPN-Lyon, Villeurbanne, France}}
\fntext[f36]{Now at: Centre de Calcul IN2P3, Lyon, France}
\fntext[f37]{Now at{ Variable Energy Cyclotron Centre, Kolkata, India}}
\fntext[f38]{Also at{ Dipartimento di Fisica dell'Universit\`{a} and Sezione INFN, Padova, Italy}}
\fntext[f39]{Also at{ Sezione INFN, Bologna, Italy}}
\fntext[f40]{Also at Dipartimento di Fisica dell\'{ }Universit\`{a}, Udine, Italy}
\fntext[f41]{Also at Wroc{\l}aw University, Wroc{\l}aw, Poland}
\fntext[f42]{Now at Dipartimento di Fisica dell'Universit\`{a} and Sezione INFN, Padova, Italy}

\address[1]{Dipartimento di Scienze e Tecnologie Avanzate dell'Universit\`{a} del Piemonte Orientale and Gruppo Collegato INFN, Alessandria, Italy}

\address[2]{Department of Physics Aligarh Muslim University, Aligarh, India}

\address[3]{Nikhef, National Institute for Subatomic Physics, Amsterdam, Netherlands}

\address[4]{Physics Department, University of Athens, Athens, Greece}

\address[5]{Dipartimento Interateneo di Fisica `M.~Merlin' and Sezione INFN, Bari, Italy}

\address[6]{Sezione INFN, Bari, Italy}

\address[7]{China Institute of Atomic Energy, Beijing, China}

\address[8]{Department of Physics and Technology, University of Bergen, Bergen, Norway}

\address[9]{Faculty of Engineering, Bergen University College, Bergen, Norway}

\address[10]{Lawrence Berkeley National Laboratory, Berkeley, CA, United States}

\address[11]{Institute of Physics, Bhubaneswar, India}

\address[12]{School of Physics and Astronomy, University of Birmingham, Birmingham, United Kingdom}

\address[13]{Dipartimento di Fisica dell'Universit\`{a} and Sezione INFN, Bologna, Italy}

\address[14]{Sezione INFN, Bologna, Italy}

\address[15]{Faculty of Mathematics, Physics and Informatics, Comenius University, Bratislava, Slovakia}

\address[16]{Institute of Space Sciences (ISS), Bucharest, Romania}

\address[17]{National Institute for Physics and Nuclear Engineering, Bucharest, Romania}

\address[18]{KFKI Research Institute for Particle and Nuclear Physics, Hungarian Academy of Sciences, Budapest, Hungary}

\address[19]{Dipartimento di Fisica dell'Universit\`{a} and Sezione INFN, Cagliari, Italy}

\address[20]{Sezione INFN, Cagliari, Italy}

\address[21]{Universidade Estadual de Campinas (UNICAMP), Campinas, Brazil}

\address[22]{Physics Department, University of Cape Town, iThemba Laboratories, Cape Town, South Africa}

\address[23]{Dipartimento di Fisica e Astronomia dell'Universit\`{a} and Sezione INFN, Catania, Italy}

\address[24]{Sezione INFN, Catania, Italy}

\address[25]{Physics Department, Panjab University, Chandigarh, India}

\address[26]{Laboratoire de Physique Corpusculaire (LPC), Clermont Universit\'{e}, Universit\'{e} Blaise Pascal, CNRS--IN2P3, Clermont-Ferrand, France}

\address[27]{Department of Physics, Ohio State University, Columbus, OH, United States}

\address[28]{Niels Bohr Institute, University of Copenhagen, Copenhagen, Denmark}

\address[29]{The Henryk Niewodniczanski Institute of Nuclear Physics, Polish Academy of Sciences, Cracow, Poland}

\address[30]{Universidad Aut\'{o}noma de Sinaloa, Culiac\'{a}n, Mexico}

\address[31]{Research Division and ExtreMe Matter Institute EMMI, GSI Helmholtzzentrum f\"{u}r Schwerionenforschung, Darmstadt, Germany}

\address[32]{Institut f\"{u}r Kernphysik, Technische Universit\"{a}t Darmstadt, Darmstadt, Germany}

\address[33]{Wayne State University, Detroit, MI, United States}

\address[34]{Joint Institute for Nuclear Research (JINR), Dubna, Russia}

\address[35]{Frankfurt Institute for Advanced Studies, Johann Wolfgang Goethe-Universit\"{a}t Frankfurt, Frankfurt, Germany}

\address[36]{Institut f\"{u}r Kernphysik, Johann Wolfgang Goethe-Universit\"{a}t Frankfurt, Frankfurt, Germany}

\address[37]{Laboratori Nazionali di Frascati, INFN, Frascati, Italy}

\address[38]{Gangneung-Wonju National University, Gangneung, South Korea}

\address[39]{Petersburg Nuclear Physics Institute, Gatchina, Russia}

\address[40]{European Organization for Nuclear Research (CERN), Geneva, Switzerland}

\address[41]{Laboratoire de Physique Subatomique et de Cosmologie (LPSC), Universit\'{e} Joseph Fourier, CNRS-IN2P3, Institut Polytechnique de Grenoble, Grenoble, France}

\address[42]{Centro de Aplicaciones Tecnol\'{o}gicas y Desarrollo Nuclear (CEADEN), Havana, Cuba}

\address[43]{Kirchhoff-Institut f\"{u}r Physik, Ruprecht-Karls-Universit\"{a}t Heidelberg, Heidelberg, Germany}

\address[44]{Physikalisches Institut, Ruprecht-Karls-Universit\"{a}t Heidelberg, Heidelberg, Germany}

\address[45]{Hiroshima University, Hiroshima, Japan}

\address[46]{University of Houston, Houston, TX, United States}

\address[47]{Physics Department, University of Rajasthan, Jaipur, India}

\address[48]{Physics Department, University of Jammu, Jammu, India}

\address[49]{Helsinki Institute of Physics (HIP) and University of Jyv\"{a}skyl\"{a}, Jyv\"{a}skyl\"{a}, Finland}

\address[50]{Bogolyubov Institute for Theoretical Physics, Kiev, Ukraine}

\address[51]{University of Tennessee, Knoxville, TN, United States}

\address[52]{Saha Institute of Nuclear Physics, Kolkata, India}

\address[53]{Variable Energy Cyclotron Centre, Kolkata, India}

\address[54]{Fachhochschule K\"{o}ln, K\"{o}ln, Germany}

\address[55]{Faculty of Science, P.J.~\v{S}af\'{a}rik University, Ko\v{s}ice, Slovakia}

\address[56]{Institute of Experimental Physics, Slovak Academy of Sciences, Ko\v{s}ice, Slovakia}

\address[57]{Laboratori Nazionali di Legnaro, INFN, Legnaro, Italy}

\address[58]{Secci\'{o}n F\'{\i}sica, Departamento de Ciencias, Pontificia Universidad Cat\'{o}lica del Per\'{u}, Lima, Peru}

\address[59]{Lawrence Livermore National Laboratory, Livermore, CA, United States}

\address[60]{Division of Experimental High Energy Physics, University of Lund, Lund, Sweden}

\address[61]{Centro de Investigaciones Energ\'{e}ticas Medioambientales y Tecnol\'{o}gicas (CIEMAT), Madrid, Spain}

\address[62]{Instituto de Ciencias Nucleares, Universidad Nacional Aut\'{o}noma de M\'{e}xico, Mexico City, Mexico}

\address[63]{Instituto de F\'{\i}sica, Universidad Nacional Aut\'{o}noma de M\'{e}xico, Mexico City, Mexico}

\address[64]{Centro de Investigaci\'{o}n y de Estudios Avanzados (CINVESTAV), Mexico City and M\'{e}rida, Mexico}

\address[65]{Institute for Nuclear Research, Academy of Sciences, Moscow, Russia}

\address[66]{Institute for Theoretical and Experimental Physics, Moscow, Russia}

\address[67]{Moscow Engineering Physics Institute, Moscow, Russia}

\address[68]{Russian Research Centre Kurchatov Institute, Moscow, Russia}

\address[69]{Indian Institute of Technology, Mumbai, India}

\address[70]{Institut f\"{u}r Kernphysik, Westf\"{a}lische Wilhelms-Universit\"{a}t M\"{u}nster, M\"{u}nster, Germany}

\address[71]{SUBATECH, Ecole des Mines de Nantes, Universit\'{e} de Nantes, CNRS-IN2P3, Nantes, France}

\address[72]{Yale University, New Haven, CT, United States}

\address[73]{Budker Institute for Nuclear Physics, Novosibirsk, Russia}

\address[74]{Oak Ridge National Laboratory, Oak Ridge, TN, United States}

\address[75]{Physics Department, Creighton University, Omaha, NE, United States}

\address[76]{Institut de Physique Nucl\'{e}aire d'Orsay (IPNO), Universit\'{e} Paris-Sud, CNRS-IN2P3, Orsay, France}

\address[77]{Department of Physics, University of Oslo, Oslo, Norway}

\address[78]{Dipartimento di Fisica dell'Universit\`{a} and Sezione INFN, Padova, Italy}

\address[79]{Sezione INFN, Padova, Italy}

\address[80]{Faculty of Nuclear Sciences and Physical Engineering, Czech Technical University in Prague, Prague, Czech Republic}

\address[81]{Institute of Physics, Academy of Sciences of the Czech Republic, Prague, Czech Republic}

\address[82]{Institute for High Energy Physics, Protvino, Russia}

\address[83]{Benem\'{e}rita Universidad Aut\'{o}noma de Puebla, Puebla, Mexico}

\address[84]{Pusan National University, Pusan, South Korea}

\address[85]{Nuclear Physics Institute, Academy of Sciences of the Czech Republic, \v{R}e\v{z} u Prahy, Czech Republic}

\address[86]{Dipartimento di Fisica dell'Universit\`{a} `La Sapienza' and Sezione INFN, Rome, Italy}

\address[87]{Sezione INFN, Rome, Italy}

\address[88]{Commissariat \`{a} l'Energie Atomique, IRFU, Saclay, France}

\address[89]{Dipartimento di Fisica `E.R.~Caianiello' dell'Universit\`{a} and Sezione INFN, Salerno, Italy}

\address[90]{California Polytechnic State University, San Luis Obispo, CA, United States}

\address[91]{Departamento de F\'{\i}sica de Part\'{\i}culas and IGFAE, Universidad de Santiago de Compostela, Santiago de Compostela, Spain}

\address[92]{Universidade de S\~{a}o Paulo (USP), S\~{a}o Paulo, Brazil}

\address[93]{Russian Federal Nuclear Center (VNIIEF), Sarov, Russia}

\address[94]{Department of Physics, Sejong University, Seoul, South Korea}

\address[95]{Yonsei University, Seoul, South Korea}

\address[96]{Technical University of Split FESB, Split, Croatia}

\address[97]{V.~Fock Institute for Physics, St. Petersburg State University, St. Petersburg, Russia}

\address[98]{Institut Pluridisciplinaire Hubert Curien (IPHC), Universit\'{e} de Strasbourg, CNRS-IN2P3, Strasbourg, France}

\address[99]{University of Tokyo, Tokyo, Japan}

\address[100]{Dipartimento di Fisica dell'Universit\`{a} and Sezione INFN, Trieste, Italy}

\address[101]{Sezione INFN, Trieste, Italy}

\address[102]{University of Tsukuba, Tsukuba, Japan}

\address[103]{Dipartimento di Fisica Sperimentale dell'Universit\`{a} and Sezione INFN, Turin, Italy}

\address[104]{Sezione INFN, Turin, Italy}

\address[105]{Nikhef, National Institute for Subatomic Physics and Institute for Subatomic Physics of Utrecht University, Utrecht, Netherlands}

\address[106]{Universit\'{e} de Lyon, Universit\'{e} Lyon 1, CNRS/IN2P3, IPN-Lyon, Villeurbanne, France}

\address[107]{Soltan Institute for Nuclear Studies, Warsaw, Poland}

\address[108]{Warsaw University of Technology, Warsaw, Poland}

\address[109]{Purdue University, West Lafayette, IN, United States}

\address[110]{Zentrum f\"{u}r Technologietransfer und Telekommunikation (ZTT), Fachhochschule Worms, Worms, Germany}

\address[111]{Hua-Zhong Normal University, Wuhan, China}

\address[112]{Yerevan Physics Institute, Yerevan, Armenia}

\address[113]{Rudjer Bo\v{s}kovi\'{c} Institute, Zagreb, Croatia}

\begin{abstract}
The inclusive charged particle 
transverse momentum distribution 
is measured in proton-proton collisions at $\sqrt{s} = 900$~GeV at 
the LHC using 
the ALICE detector. 
The measurement is performed in the central pseudorapidity 
region $(|\eta|<0.8)$
over the transverse momentum range $0.15<p_T<10$~GeV/$c$. The correlation 
between transverse momentum and particle multiplicity is also studied. 
Results are presented for inelastic (INEL) and non-single-diffractive (NSD)
events. 
The average transverse momentum for $|\eta|<0.8$ is 
\mpt$_{\rm INEL}=0.483\pm0.001$~(stat.)~$\pm0.007$~(syst.)~GeV/$c$ and  
\mpt$_{\rm NSD}=0.489\pm0.001$~(stat.)~$\pm0.007$~(syst.)~GeV/$c$,
respectively.
The data exhibit a slightly
larger \mpt 
than measurements in 
wider pseudorapidity
intervals. The results are compared to 
simulations with the Monte Carlo event generators PYTHIA and PHOJET. 
\end{abstract}



\end{frontmatter}

\begin{multicols}{2}
\section{Introduction}
\label{}

The precise measurement of the transverse momentum spectrum of charged
particles produced in proton collisions in the energy range of the 
Large Hadron Collider (LHC)~\cite{lhc} offers unique information 
about soft and hard interactions.
Perturbative Quantum Chromodynamics (pQCD) is a 
framework for the quantitative
description of parton-parton interactions at large momentum transfers, 
i.e.~hard scattering 
processes.
However, a significant fraction of the particles produced in $pp$ collisions 
do not originate from hard interactions, even at LHC energies.
In contrast to hard processes, the description of particle production
in soft interactions is not well-established within QCD. 
Current models of hadron-hadron collisions at high energies, such
as the event generators PYTHIA~\cite{pyth1} and PHOJET~\cite{phojet},
combine perturbative QCD for the description
of hard parton interactions with phenomenological approaches to model
the soft component of the produced particle spectrum. 
Data on charged particle production in hadron-hadron
collisions will have to be used to tune these models 
before they can 
provide a detailed description
of the existing measurements 
and predictions 
for particle production characteristics in $pp$ collisions
at the highest LHC energies.
These data include 
the measurement of multiplicity, pseudorapidity ($\eta$) and
transverse momentum $(p_T)$ distributions of charged particles and 
correlations, such as the dependence of the average
transverse momentum, $\left<p_T\right>$, on the charged particle multiplicity.

The charged particle pseudorapidity densities and multiplicity
distributions in $pp$ collisions 
at $\sqrt{s}=$~0.9, 2.36 and 7~TeV were presented
in recent publications by the ALICE 
collaboration~\cite{paper1,paper2,paper7tev}.
In this letter, we present a measurement in  
$pp$~collisions at $\sqrt{s} = 900$~GeV of the transverse 
momentum spectrum of primary charged particles and the 
correlation between \mpt and the charged particle multiplicity. 
Primary particles include particles
produced in the collision or their decay products,
except those from weak decays of strange hadrons.
The measurement is performed in the central rapidity region 
$(|\eta|<0.8)$ and 
covers a $p_T$ range $ 0.15 < p_T < 10$~GeV/$c$,
where both hard and soft processes are expected to contribute
to particle production.
The data from the ALICE experiment presented in this 
letter serve as a baseline 
for future studies of $pp$ collisions at higher 
LHC energies and 
particle 
production in heavy-ion
collisions~\cite{alice-ppr}.

\section{Experiment and data collection}
The data were collected with the 
ALICE detector~\cite{alice-det} during the startup phase of the LHC 
in December 2009.
The ALICE detector, designed to cope with 
high track densities in heavy-ion collisions, provides
excellent track reconstruction and particle identification
capabilities. 
This also makes the detector well-suited to detailed studies of 
global characteristics of $pp$ interactions~\cite{alice-ppr}. 

In this analysis of the first $pp$ collisions at $\sqrt{s} = 900$~GeV, 
charged particle
tracking and momentum reconstruction are based on data recorded with the 
Time Projection Chamber (TPC) and the Inner Tracking System (ITS), 
both located in the central barrel of ALICE.
The detectors in the central barrel are operated inside a large solenoidal 
magnet providing a uniform 0.5~T field.

The ALICE TPC~\cite{tpc-nim} 
is a large cylindrical drift detector
with a central high voltage membrane maintained at $-100$~kV 
and two readout planes at the
end-caps. The active volume is limited to $85<r<247$~cm  and
$-250 < z < 250$~cm in the radial and longitudinal directions respectively. 
The material budget between the interaction point and the 
active volume of the TPC corresponds to 11\% of a radiation length,
averaged in $|\eta| < 0.8$.
The central membrane at $z=0$
divides the nearly 90~m$^3$ active volume into two halves.
The homogeneous drift field of 400~V/cm in the Ne-CO$_2$-N$_2$ (85.7\%-9.5\%-4.8\%)
gas mixture leads to a maximum drift time of 94~$\mu$s.
Ionization electrons produced by charged particles traversing the 
TPC drift towards the readout end-caps composed
of 72 multi-wire proportional chambers with cathode pad readout.
The typical gas gain is $10^4$. 
Signals induced on the segmented
cathode planes, comprising a total of 558k readout pads,
are transformed into differential semi-gaussian signals by 
a charge-sensitive shaping amplifier (PASA).
This is followed by the ALICE TPC ReadOut (ALTRO) chip,
which employs a 10 bit ADC at 10 MHz 
sampling rate and four digital filtering circuits. These filters also
perform tail 
cancellation and baseline restoration. They are optimized for precise position
and d$E$/d$x$ measurements in the high track density environment of heavy-ion
collisions. To ensure optimal drift and charge transport properties,
the TPC was operated with an overall temperature uniformity of 
$\Delta T \approx 60$~mK (r.m.s.).
The oxygen contamination was less than $5$~ppm.

The ITS is composed of high resolution 
silicon tracking detectors, arranged in six cylindrical layers  
at radial distances to the beam line from 3.9 to 43 cm. 
Three different technologies are employed.

For the two innermost layers Silicon Pixel Detectors (SPD) are used,
covering the 
pseudorapidity ranges $|\eta| < 2$ and $|\eta| < 1.4$, respectively. 
A total of 9.8 million
$50 \times 425$~$\mu$m$^2$ pixels enable the reconstruction of the
primary event vertex and the track impact parameters with high precision. 
The SPD was also included in the trigger scheme for data collection.

The SPD is followed by two Silicon Drift Detector (SDD) layers
with a total of 133k readout channels, sampling the drift
time information at a frequency of 20~MHz. 
The SDD are operated with a drift field of 500~V/cm, resulting in
a drift speed of about 6.5~$\mu$m/ns and in a maximum drift time
of about 5.3~$\mu$s.

The two outermost Silicon Strip Detector (SSD) layers
consist of double-sided silicon micro-strip sensors with $95~\mu$m pitch,
comprising a total of 2.6 million readout channels. 
Strips of the two sensor sides form a stereo
angle of 35~mrad, providing two-dimensional hit reconstruction.

The design spatial resolutions of the ITS sub-detectors 
($\sigma_{r\phi} \times \sigma_{z}$) are: $12\times 100~\mu{\rm m}^2$ for SPD,
$35\times 25~\mu{\rm m}^2$ for SDD, and $20 \times 830~\mu{\rm m}^2$ for SSD. 
The SPD and SSD detectors were aligned using survey measurements, 
cosmic muon data~\cite{alignment} and collision data to an estimated 
accuracy of 10~$\mu$m for the SPD and 15~$\mu$m for the SSD.
No alignment corrections are applied to the positions of the 
SDD modules, for which 
calibration and alignment are in progress. The estimated misalignment
of the SDD modules is about 100~$\mu$m.
The TPC and ITS are aligned relative to each other to the level of a 
few hundred micrometers using cosmic-ray and $pp$ data
by comparing pairs of track segments independently reconstructed 
in the two detectors.

The two forward scintillator hodoscopes (VZERO) are included in the
trigger. 
Each detector is segmented into 32 scintillator counters which 
are arranged in four rings around the beam pipe. They are located at distances
$z=3.3$~m and $z=-0.9$~m from the nominal interaction point
and cover the pseudorapidity ranges: $2.8 < \eta <5.1$ and $-3.7 < \eta <-1.7$
respectively. 
The time resolution of about 1~ns of the VZERO hodoscope 
also allows for a discrimination against beam-gas interactions.

During the startup phase of the LHC in 2009, 
four proton bunches per beam were circulating in the
LHC with two pairs of bunches crossing at the 
ALICE intersection region and protons colliding at $\sqrt{s}=900$~GeV. 
The detector readout was triggered using the LHC bunch-crossing signals
in coincidence with signals from the two upstream beam pick-up counters
and a minimum-bias interaction trigger requiring a 
signal in at 
least one of the SPD pixels or one of the VZERO counters~\cite{paper1,paper2}. 
Events with only one bunch or 
no bunches 
passing through ALICE were also recorded
to study beam related and random background.

\section{Data analysis}

The total inelastic $pp$ cross section is commonly subdivided into  
contributions from diffractive and non-diffractive processes.
To facilitate comparison with existing measurements, we perform our analysis
for two classes of events: inelastic (INEL) and non-single-diffractive (NSD) 
$pp$ collisions. 

In this analysis, 
$3.44\times10^5$ triggered $pp$ events at 
$\sqrt{s}=900$~GeV are analyzed. To remove beam related background 
events, an offline event selection 
based on the VZERO timing signal and the correlation between the number
of hits and tracklets in the SPD 
is applied as in~\cite{paper2},
reducing the sample to $2.67\times10^5$ events. 
This event selection is refered to as \mbor~\cite{paper2}.

For the INEL analysis we use the event sample
selected with the MB$_{\rm OR}$  condition. A subset
of these events ($2.15\times10^5$) is used for the NSD analysis, 
selected offline by requiring a 
coincidence between the two VZERO detectors (the \mband selection). 
This condition 
suppresses a significant fraction of the single-diffractive events and hence
reduces the systematic errors related to model dependent 
corrections~\cite{paper2}.

The fractions of the different process types contributing to the 
selected event 
samples are estimated by a Monte Carlo simulation, implementing
a description of the ALICE detector response~\cite{aliroot}
to $pp$ collisions at $\sqrt{s}=900$~GeV from the PYTHIA 
event generator version 6.4.21 tune D6T (109)~\cite{109-tune}. 
The process fractions of single-diffractive (SD) and double-diffractive (DD)
events are scaled in Monte Carlo to match the cross sections in 
$p\bar{p}$ at $\sqrt{s}=900$~GeV measured by UA5~\cite{ua5-diffraction}.
The selection efficiency for INEL events using \mbor
and NSD events using \mband is approximately $96$\% and $93$\%, 
respectively~\cite{paper2}. 

Charged particle tracks are reconstructed using information from
the TPC and ITS detector systems.
Signals on adjacent pads in the TPC are connected to particle tracks
by employing a Kalman filter algorithm. 
The TPC tracks
are extrapolated to the ITS and matching hits in the ITS detector layers 
are assigned to the track.
In order to maximize the hit matching efficiency and avoid possible 
biases of the track parameters due to the non-uniform degree of alignment
of the ITS sub-detectors, the space point uncertainties of the ITS hits 
are set to $100~\mu$m for SPD and 1~mm for both SDD and SSD. 

The event vertex is reconstructed using the combined track information
from TPC and ITS. 
The tracks are extrapolated
to the intersection region and the position of the event vertex is fitted,
using the measured average intersection profile as a constraint. 
The profile of the intersection region is determined on a run-by-run basis
in a first pass through the data using the mean and the spread of the
distribution of the reconstructed vertices.
The event vertex distribution is found to be Gaussian with standard deviations 
of approximately $210~\mu$m,
$250~\mu$m, and $4.1~$cm, along $x$, $y$ (transverse to the beam-axis) 
and $z$ respectively.
For events where only one track is found, the vertex is determined
from the point of closest approach of the track to the beam axis.
If no track
is found in the TPC, the event vertex reconstruction is based on 
tracklets built by associating pairs of hits of the two innermost 
ITS layers (SPD).
An event with a reconstructed vertex position $z_v$ is accepted if 
$|z_v - z_0|<10$~cm, corresponding to about 2.5 standard deviations of
the reconstructed event vertex distribution centered at $z_0$~\cite{paper2}.

The vertex position resolution depends on the event multiplicity. It  
can be parametrized as $540~\mu$m$/(N_{\rm SPD})^{0.45}$ in $x$ and $y$, 
and $550~\mu$m/$(N_{\rm SPD})^{0.6}$
in $z$, where $N_{\rm SPD}$ corresponds to the number of SPD 
tracklets. This resolution is consistent with 
Monte Carlo simulations.
The probability of multiple interactions in the same bunch crossing (pile-up) 
in the present data set is $10^{-4}$ and therefore neglected.

The fraction of selected events in the \mbor (\mband) sample
where an event vertex
is successfully reconstructed is 80\% (92\%), resulting in 
a sample of $2.13\times10^5$ INEL ($1.98\times10^5$ NSD) 
events used in the present analysis.
Events, where no vertex is found, are included when normalizing the results.
In order to understand and to subtract possible beam-induced background, 
the detector was also triggered on bunches coming from either side
of the interaction region,
but not colliding with another bunch. 
From the study of these events we estimate that 21\% of the triggered
\mbor and \mband
events without a reconstructed event vertex or with zero selected tracks are 
background, and the number of events used for normalization of
the final results is corrected accordingly. 
The estimated contribution from beam-induced background events to the 
event sample, 
where a vertex was found, is negligible.
From the analysis of
empty bunch crossing events the random contribution from 
cosmics and noise triggers is also found to be negligible.

To study the transverse momentum spectrum, charged 
particle tracks are selected in the 
pseudorapidity range $|\eta|<0.8$. 
In this range, tracks in the TPC can be reconstructed with maximal length,
and there are minimal efficiency losses due to detector boundaries.
Additional quality requirements are applied to ensure high tracking 
resolution and low secondary and fake track contamination. 
A track is accepted if it has at least 70 out of the 
maximum of 159 space points in the TPC,
and the $\chi^2$ per space point used for the momentum fit is less than~4. 
Additionally, at least two hits in the ITS must be associated 
with the track, and at least one has to be in either of the two innermost
layers, i.e., in the SPD.
The average number of associated hits per track in the six ITS layers
is 4.7, mainly determined by the fraction of inactive channels, and is well
reproduced in Monte Carlo simulations.
Tracks with $p_T < 0.15$~GeV/$c$ are
excluded because their reconstruction efficiency drops below~50\%.
Tracks are also rejected as not associated to the primary vertex
if their distance of closest approach to the reconstructed event 
vertex in the plane perpendicular to the beam axis, $d_0$,
satisfies $d_0 > 0.35{\rm ~mm}+0.42{\rm ~mm}\times p_T^{-0.9}$, with $p_T$ in
GeV/$c$.
This cut corresponds to about seven standard deviations of the $p_T$ dependent
transverse impact parameter resolution for primary tracks passing the
above selection. It is tuned 
to select primary charged particles with high
efficiency and to minimize the contributions from weak decays, 
conversions and secondary hadronic interactions in the detector 
material. The accepted number of charged particles per event which 
fulfill these conditions is called $n_{\rm acc}$.

With this selection, the reconstruction efficiency for primary 
charged particles and the remaining contamination from secondaries
as a function of $p_T$ are
estimated by Monte Carlo simulation 
using PYTHIA, combined with detector simulation and 
event reconstruction. 
The procedure estimates losses due to tracking inefficiency,
charged particles escaping detection due to weak decay, absorption
and secondary interaction in the detector. 
The inefficiencies of the event
selection and of the event vertex reconstruction are accounted for.
The latter two affect mostly low-multiplicity events, which imposes
a bias on the uncorrected $p_T$ spectrum due to the correlation 
between multiplicity and average momentum.

\begin{figurehere}
\begin{center}
\includegraphics[width=8.0cm]{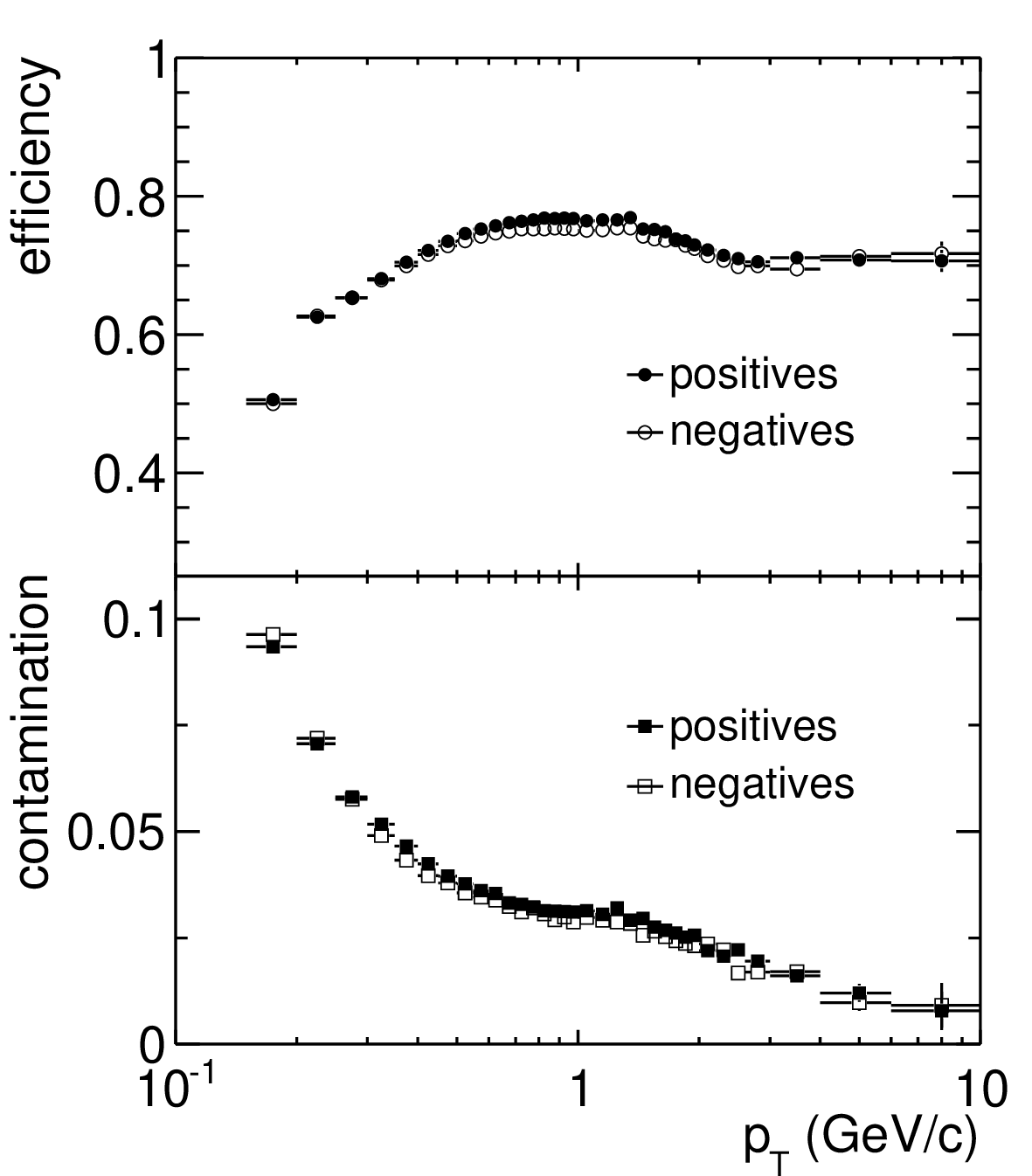}
\caption{\label{eff-sec} 
Charged-particle track reconstruction efficiency for primary particles (top)
and contamination from secondary particles (bottom), 
for positively and negatively charged particles in $|\eta|<0.8$ as
a function of $p_T$. The tracking efficiency is normalized to the
number of generated primary particles using PYTHIA. The contamination
from secondary tracks was scaled in Monte Carlo to match the 
measured $d_0$ distributions
(see text).} 
\end{center}
\end{figurehere}

The primary charged particle track reconstruction efficiency in the
region $|\eta|<0.8$ reaches 75\% at $p_T \sim 1$~GeV/$c$,
as shown in Fig.~\ref{eff-sec}. 
The slight decrease
of efficiency observed for $p_T>1.5$~GeV/$c$ is a consequence
of the projective segmentation of the readout plane in azimuth,
causing stiff tracks to remain undetected if they fall between
two adjacent TPC readout sectors. 
For $p_T<0.6$~GeV/$c$, the reconstruction
efficiency decreases and reaches 50\% at 0.15~GeV/$c$.
The losses at low $p_T$ are mainly due to energy loss in the detector
material and to the track bending in the magnetic field. 
No significant dependence of the track reconstruction efficiency on the 
track density is observed in simulations for charged particle multiplicities
relevant for this analysis.
The contamination from secondary particles such as charged particles
from weak decays, electrons from photon conversions, and products from 
secondary interactions in the detector material is 
also shown in Fig.~\ref{eff-sec}.
It has a maximum of 9\% at the lowest $p_T$ and drops below 3\% for
$p_T> 1$~GeV/$c$.
A  comparison of the $d_0$ distributions
of data and Monte Carlo tracks indicates that the Monte Carlo simulation
using PYTHIA underestimates
the particle yield from secondaries by 0-50\%, depending on $p_T$. This is
consistent with the fact that PYTHIA underestimates the strangeness yield by 
a similar amount, when compared to previous results in $pp$ and $p\bar{p}$ 
collisions~\cite{strangeness,strange-star}. 
For the final corrections to the data we scale accordingly
the contamination level obtained with PYTHIA, resulting in an
additional  0-1.5\% decrease of the primary particle yields. 
The uncertainty in the strangeness yield is taken into account in the 
evaluation of the overall systematic uncertainties, as discussed below.

The reconstruction efficiency and contamination are converted
to $p_T$ dependent correction factors used to correct the raw $p_T$ spectrum.
We note that efficiency and secondary contamination are slightly 
different for positively and negatively charged particles, mainly due to the larger absorption 
of negatively charged particles and isospin effects in secondary interactions.

The charged particle transverse momenta are measured
in the TPC, taking into account energy loss 
based on the PID hypothesis from TPC d$E$/d$x$ and
the material budget in front of the TPC. 
The material budget is studied
via the 
measurement of electron-positron pairs in the TPC from photon conversions.
The radial distribution
of the reconstructed photon conversion points is compared to Monte Carlo
simulations. The sum of all positive and negative deviations is $+4.7$\% 
and $-7.2$\%, respectively.
The remaining material budget uncertainty enters into the final
systematic uncertainties. In this analysis, we use the measurement of
the momentum at the event vertex.

At the present level of calibration, the transverse momentum resolution
achieved in the TPC is given by
$(\sigma(p_T)/p_T)^2 = (0.01)^2+(0.007 \cdot p_T)^2$, 
with $p_T$ in GeV/$c$. 
The transverse momentum resolution for $p_T>1$~GeV/$c$ is measured in
cosmic muon events by comparing the muon momenta reconstructed
in the upper and lower halves of the TPC. For $p_T<1$~GeV/$c$, the
Monte Carlo estimate of $\sigma(p_T)/p_T\approx 1\%$ is cross-checked 
using the measured K$^0_s$ invariant mass distribution.
A Monte Carlo based correction is applied to the $p_T$ spectra
to account for the finite momentum resolution. 
The correction increases with $p_T$ and reaches 1.2\% at 10~GeV/$c$.
 
The calibration of the absolute momentum scale  
is verified employing the invariant mass spectra
of $\Lambda$, $\bar{\Lambda}$, K$^0_s$ and $\phi$. 
The reconstructed peak positions agree with their PDG
values within 0.3~MeV/$c^2$.
As a cross-check, the $q/p_T$ distributions
of particles with charge $q$ in data and Monte Carlo simulation
are compared and the 
symmetry of the minimum around $q/p_T=0$ is studied.
Based on these studies, we estimate   
an upper limit on the systematic 
uncertainty of the momentum
scale of $|\Delta (p_T)/p_T| < 0.003$.
Within the $p_T$ reach of this study, the effect of the
momentum scale uncertainty on the final spectra
is found to be negligible.

For the normalization of the transverse momentum spectra to the 
number of events, multiplicity dependent correction factors are 
derived from the event selection 
and vertex reconstruction efficiencies for INEL and NSD events,
evaluated with the PYTHIA Monte Carlo event generator.

The fully corrected $p_T$ spectra 
are fitted by the modified Hagedorn function~\cite{hagedorn}
\begin{equation}
\frac{1}{2\pi p_T}\frac{{\rm d}^2N_{\rm ch}}{{\rm d}\eta~{\rm d}p_T}\propto
\frac{p_T}{m_T} \left(1+\frac{p_T}{p_{T,0}}\right)^{-b}.
\label{param}
\end{equation}
For the transverse mass 
$m_T = \sqrt{m_{\pi}^2+p_T^2}$, the pion mass is assumed for all 
tracks.
At small $p_T$, the term $\left(1+\frac{p_T}{p_{T,0}}\right)^{-b}$ 
behaves like an exponential in 
$p_T$ with inverse
slope parameter $p_{T,0}/b$. This provides a good description of
the soft part of the spectrum, allowing for an extrapolation of the
measured data to $p_T=0$.
To assess the tail of the spectrum at $p_T>3$~GeV/$c$, a power law fit
is performed
\begin{equation}
\frac{1}{2\pi p_T}\frac{{\rm d}^2N_{\rm ch}}{{\rm d}\eta~{\rm d}p_T}\propto
p_T^{-n},
\label{powerlaw}
\end{equation}
yielding a very good description of the hard part of the
spectrum characterized by the power $n$.
  
The calculation of \mpt in all INEL and NSD events is performed
using the weighted average over the measured points in the range 
$0.15 < p_T < 10$~GeV/$c$ 
combined with the result of the Hagedorn 
fit to extrapolate to $p_T = 0$. 

In order to analyze the behaviour of
\mpt as function of multiplicity, the INEL data sample
is subdivided into bins of $n_{\rm acc}$. 
The results for \mpt are presented calculating the weighted average over
two different $p_T$ ranges, $0.15<p_T<4$~GeV/$c$ and $0.5<p_T<4$~GeV/$c$.
In addition, results are presented employing the extrapolation to $p_T = 0$ 
as described above.

To extract the correlation between \mpt and the number of primary charged 
particles ($n_{\rm ch}$) in $|\eta|<0.8$, the following weighting procedure 
is applied to account for the 
experimental resolution of the measured event multiplicities:

\begin{equation}
\left<p_T\right>(n_{\rm ch})=
\sum_{n_{\rm acc}}\left<p_T\right>(n_{\rm acc})R(n_{\rm ch},n_{\rm acc}).
\label{weighting}
\end{equation}

This method employs the normalized response matrix 
$R(n_{\rm ch},n_{\rm acc})$ from Monte Carlo simulations which contains the 
probability that an event with multiplicity $n_{\rm ch}$ is reconstructed
with multiplicity $n_{\rm acc}$. The results from this approach
are consistent with an alternative Monte Carlo based procedure, where an average multiplicity 
$\langle n_{\rm ch}\rangle$ is assigned to every measured multiplicty $n_{\rm acc}$.

\section{Systematic uncertainties}

In order to estimate the systematic uncertainties of the final $p_T$
spectra, the results of the
data analysis and of the evaluation of the corrections from Monte Carlo
simulations
are checked for stability under varying cuts and Monte
Carlo assumptions, within reasonable limits. 
In particular we studied a variation of the ratios of the most abundant
primary charged particles (p, $\pi$, K) by
$\pm30\%$ with respect to their PYTHIA values, 
the relative fractions of diffractive processes 
corresponding to their experimental errors~\cite{paper2,ua5-diffraction}, 
the TPC readout chamber alignment  ($\pm100~\mu$m),
and track and event quality cuts in the
analysis procedure. 
Particular attention was paid to the rejection efficiency of secondary
particles using the $d_0$ cut. The stability of the results under variation
of the $d_0$ cut value ($\pm3$ standard deviations with respect to the nominal
value), the secondary yield from strange hadron decays 
($\pm30$\%) and
the material budget ($\pm10\%$) was studied and the systematic
uncertainty is estimated accordingly.
Systematic uncertainties of the ITS and TPC detector efficiencies
are estimated by a comparison of the experimental ITS-TPC track
matching efficiency with the Monte Carlo one.
The systematic uncertainty of the VZERO triggering efficiency is 
studied by varying the calibration and threshold settings in the data 
and in the Monte Carlo simulation. 
The event generator dependence is determined
from a comparison of the PYTHIA results with those obtained using PHOJET.
The total systematic uncertainty on the
$p_T$ spectra derived
from this study is 3.0-7.1\% for INEL events and 
3.5-7.2\% for NSD events, in the $p_T$ range from $0.2-10$~GeV/$c$ 
(see Table~\ref{tab1}). 

Also listed in Table~\ref{tab1} are the systematic errors in \mpt
arising from these contributions. We note that only $p_T$ dependent
errors on the $p_T$ spectra contribute to the systematic error in 
$\left<p_T\right>$.
Additional systematic uncertainties in \mpt arise from the specific
choice of the fit function used for the $p_T=0$ extrapolation, and
the weighting procedure which is employed to 
derive \mpt as function
of $n_{\rm ch}$. To estimate the uncertainty in the extrapolation to $p_T=0$ 
the results are compared to those obtained from a fit of the Tsallis 
function~\cite{tsallis}, 
or by fitting the spectral shape predicted
by PYTHIA and PHOJET to our low $p_T$ data points. Based on this comparison
a systematic error of 1\% in \mpt is assigned to the $p_T=0$ extrapolation.
The weighting procedure (Eq.~\ref{weighting}) was studied using
PYTHIA and PHOJET simulations. For both models, the true \mpt dependence on
$n_{\rm ch}$ from Monte Carlo can be recovered within 3\% from the reconstructed 
dependence of \mpt on $n_{\rm acc}$ using~(\ref{weighting}). 
No significant multiplicity dependence of the 
systematic errors on \mpt is observed. The total 
systematic uncertainties on
\mpt are listed in Table~\ref{tab1}.

\section{Results and discussion}

\begin{figure*}
\centering
\includegraphics[width=8cm]{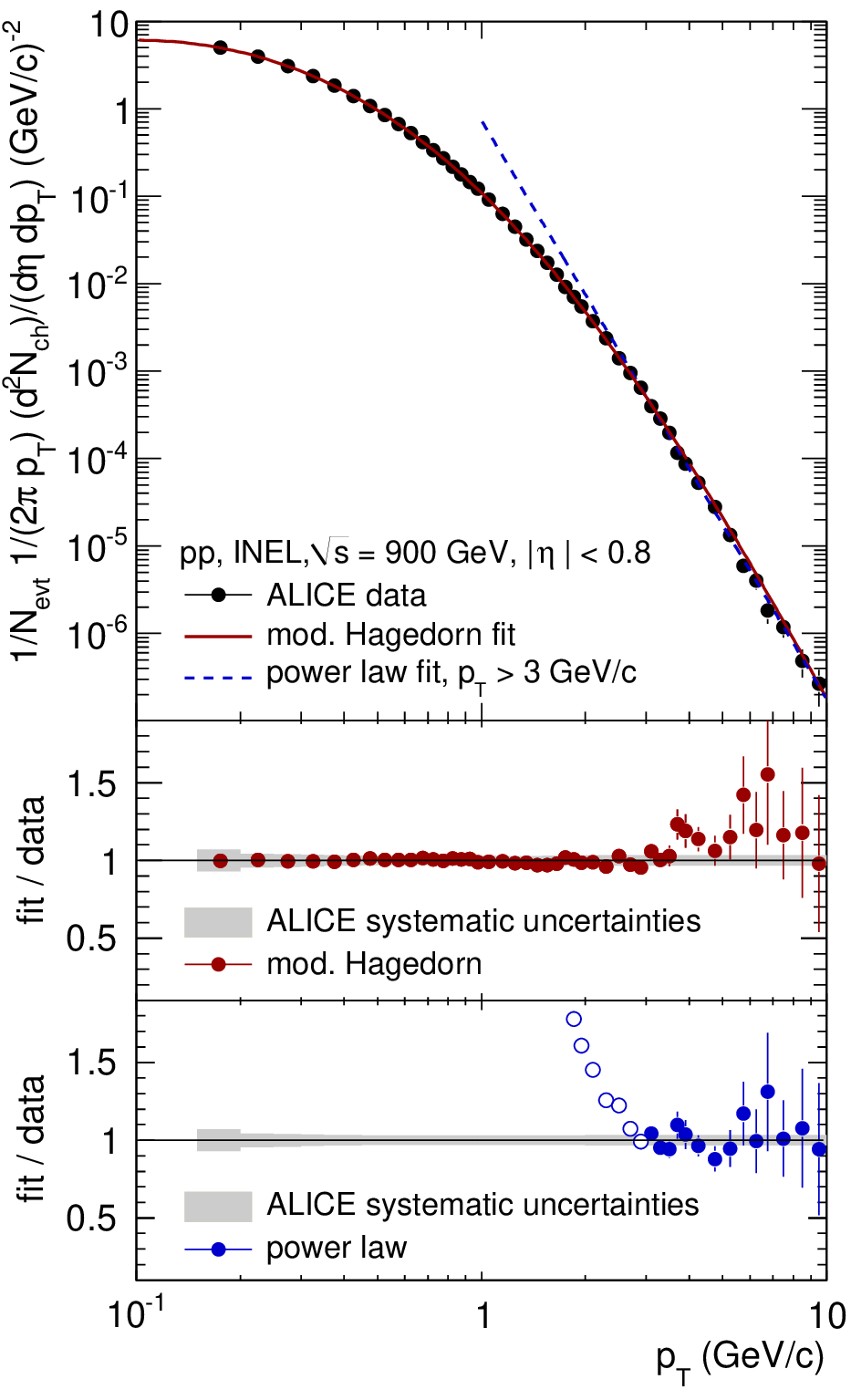}
\caption{\label{inel-fit} 
Normalized differential primary charged particle yield
in INEL $pp$ collisions at $\sqrt{s}=900$~GeV, averaged in 
$|\eta|<0.8$. The fit ranges are $0.15<p_T<10$~GeV/$c$ for the modified Hagedorn 
function (Eq.~\ref{param}) and $3<p_T<10$~GeV/$c$ for the power law (Eq.~\ref{powerlaw}).
In the lower panels, the ratios fit over data are shown. 
The open symbols indicate here data points which are not included
in the fit.
Errors bars are statistical only. Indicated as shaded
areas are the relative systematic data errors.}
\end{figure*}

The normalized differential yield in INEL $pp$ collisions
at $\sqrt{s}=900$~GeV and the fit with the parametrization 
given in Eq.~\ref{param} are shown in Fig.~\ref{inel-fit}.
The modified Hagedorn fit provides a good description of the data
for $p_T<4$~GeV/$c$. The fit parameters for INEL events are
$p_{T,0}=1.05\pm0.01$~(stat.)$\pm0.05$~(syst.)~GeV/$c$ and 
$b=7.92\pm0.03$~(stat.)$\pm0.02$~(syst.).
The average transverse momentum including the extrapolation to 
$p_T=0$ is 
\mpt$_{\rm INEL}=0.483\pm0.001$~(stat.)~$\pm0.007$(syst.)~GeV/$c$.
For NSD events we obtain 
$p_{T,0}=1.05\pm0.01$~(stat.)$\pm0.05$~(syst.)~GeV/$c$, 
$b=7.84\pm0.03$~(stat.)$\pm0.02$~(syst.)
and \mpt$_{\rm NSD}=0.489\pm0.001$~(stat.)~$\pm0.007$~(syst.)~GeV/$c$.
Restriction of the modified Hagedorn fit 
to $p_T<4$~GeV/$c$ has a negligible effect on these results.
Fig.~\ref{inel-fit} also shows the result of a 
power law fit (Eq.~\ref{powerlaw}) to the INEL data for $p_T>3$~GeV/$c$.
The power law fit provides a significantly better description of the
high $p_T$ tail of the spectrum than the modified Hagedorn parametrization.
The result of the power law fit is $n=6.63\pm0.12$~(stat.)$~\pm0.01$~(syst.)
for both INEL and NSD events. The power law shape of the high $p_T$ part
of the spectrum is suggestive of pQCD.
Estimates of differential
cross sections can be obtained using the cross sections derived from 
the measurement by UA5~\cite{ua5-diffraction} in $p\bar{p}$ 
at $\sqrt{s}=900$~GeV,
$\sigma_{\rm INEL}=50.3\pm0.4~{\rm (stat.)}\pm1~{\rm (syst.)}$~mb
and $\sigma_{\rm NSD}=42.6\pm1.4$~mb
(see also ~\cite{poghosyan}).

\begin{figure*}
\includegraphics[width=8.0cm]{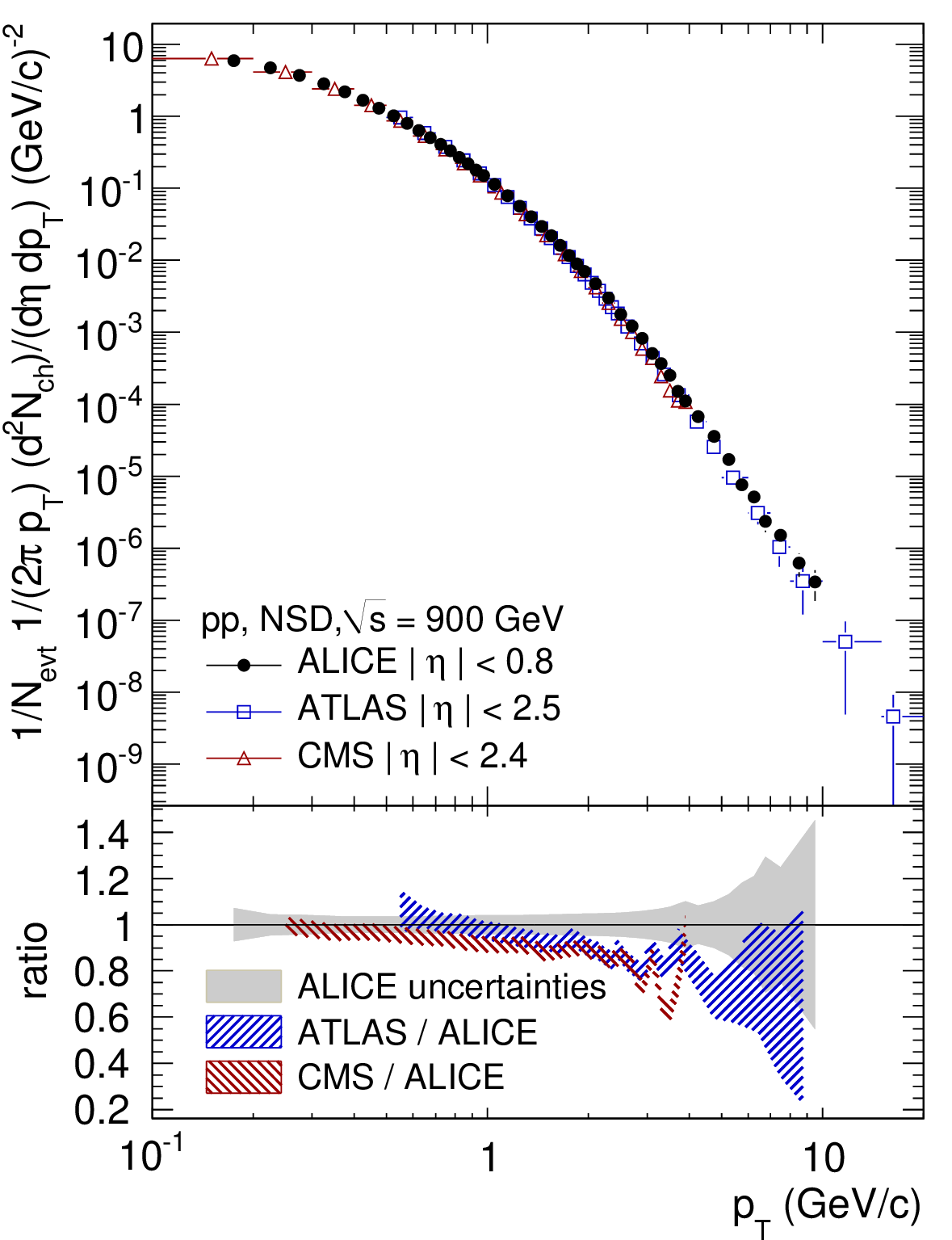}
\includegraphics[width=8.0cm]{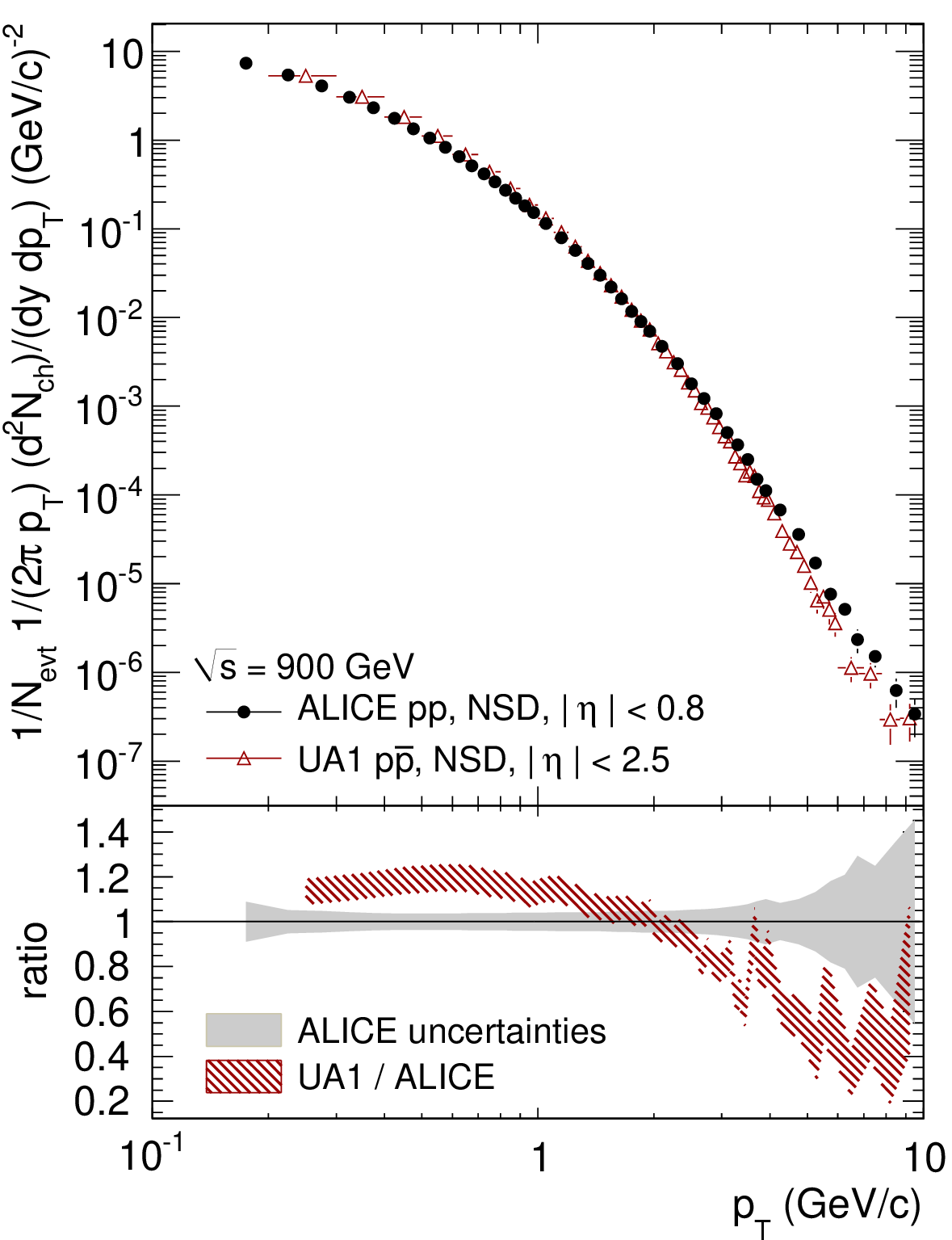}
\caption{\label{nsd-atlas-cms} 
Left panel: Normalized differential primary charged particle 
yield in NSD $pp$ collisions at $\sqrt{s}=900$~GeV, averaged in
$|\eta|<0.8$.
The ALICE data are compared to results from ATLAS
and CMS in
$pp$ at the same energy~\cite{atlas-paper1,cms-paper1}. 
Right panel:
Normalized invariant primary charged particle 
yield in NSD $pp$ collisions at $\sqrt{s}=900$~GeV, averaged in
$|\eta|<0.8$.
The ALICE data are compared to results from UA1 in
$p\bar{p}$ at the same energy~\cite{ua1}. 
For the computation of the invariant yield, it has been assumed that all particles
are pions.
The shaded areas indicate the statistical and 
systematic errors added in quadrature.}
\end{figure*}

The transverse momentum distribution for NSD events is 
shown in Fig.~\ref{nsd-atlas-cms} (left panel) together
with data recently published by ATLAS~\cite{atlas-paper1}
and CMS~\cite{cms-paper1}, 
measured in larger pseudorapidity intervals. 
Below $p_T=1$~GeV/$c$
the data agree. At higher $p_T$
the data are slightly above the other two LHC measurements.
The observation of a harder spectrum is
related to the different pseudorapidity windows (see below).

In the right panel of Fig.~\ref{nsd-atlas-cms}, the normalized invariant yield in 
NSD events is compared to
measurements of the UA1~collaboration in $p\bar{p}$ at the same 
energy~\cite{ua1}, scaled by their measured NSD cross section of 43.5~mb.
As in the previous comparison to ATLAS and CMS, the higher yield at 
large $p_T$ may be related to the
different pseudorapidity acceptances. 
The excess of the UA1 data of about 20\% at low $p_T$ 
is possibly due to the UA1 trigger condition,
which suppresses events with very low multiplicity, as pointed out 
in~\cite{atlas-paper1}.

The results for \mpt in INEL and NSD events are compared to other 
experiments~\cite{cms-paper1,ua1,isr,e735,cdf,cms-7TeV} 
in Fig.~\ref{mpt-sqrts}. 
Our results are somewhat higher than 
previous measurements
in $pp$ and $p\bar{p}$ at the same energy, but in larger pseudorapidity
windows. This is consistent with the comparison of the spectra in 
Fig.~\ref{nsd-atlas-cms}. 
A similar trend exhibiting a larger \mpt in a smaller pseudorapidity interval 
around mid-rapidity is apparent in Fig.~\ref{mpt-sqrts} at Tevatron energies.

\begin{figurehere}
\centering
\includegraphics[width=7.8cm]{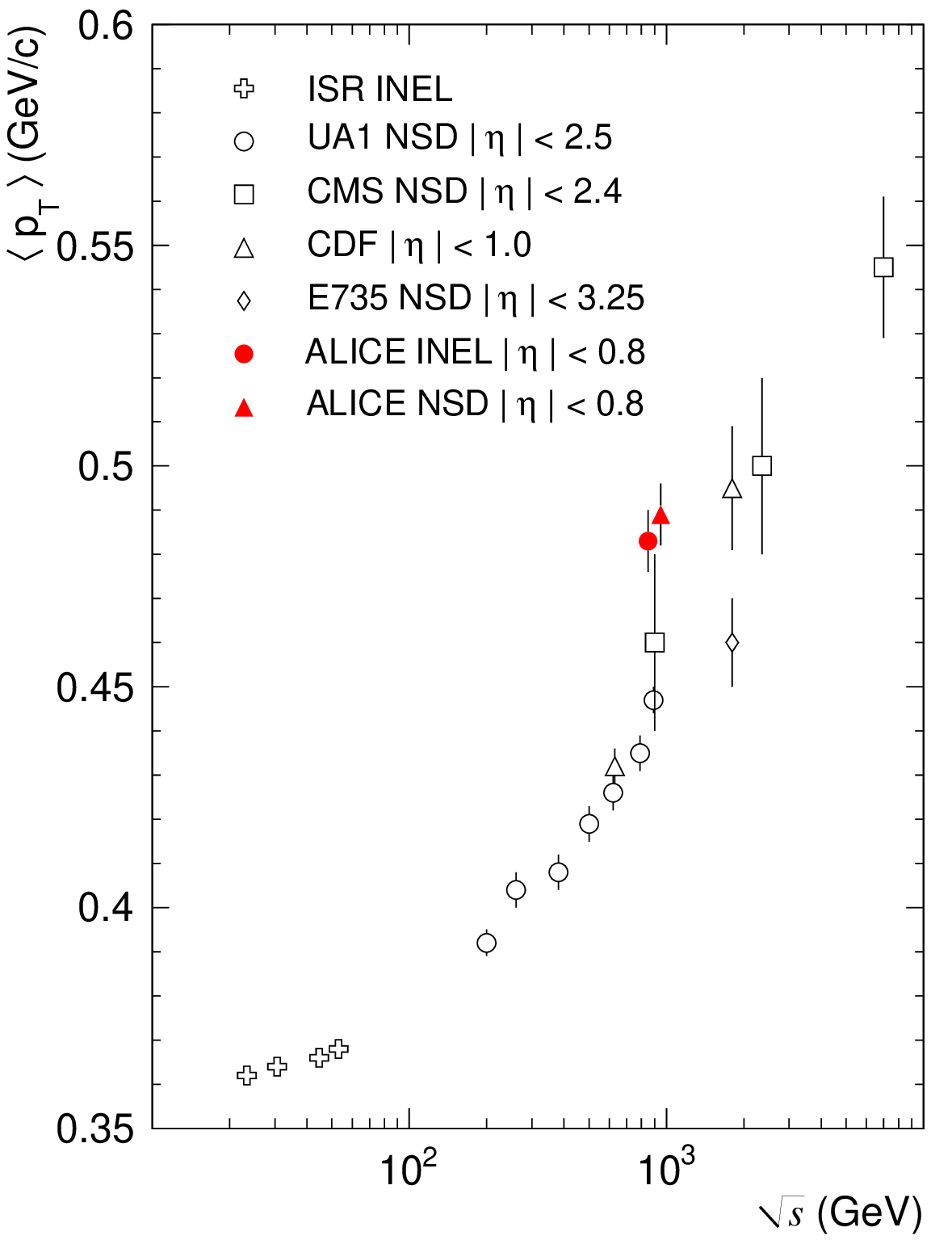}
\caption{\label{mpt-sqrts} Energy dependence of the average transverse 
momentum of 
primary charged particles in $pp$ and $p\bar{p}$ collisions. 
Data from other experiments are taken 
from~\cite{cms-paper1,ua1,isr,e735,cdf,cms-7TeV}.}
\end{figurehere}

Indeed, a decrease of \mpt
by about 2\% is found between $|\eta|<0.2$ and $0.6<|\eta|<0.8$
in a pseudorapidity dependent analysis of the present data. 
A consistent decrease of \mpt is also observed in
the CMS data, when pseudorapidity is 
increased~\cite{cms-paper1,cms-tables}.
Likewise, a decrease of \mpt by about 5\% between $|\eta|<0.8$ and 
$|\eta|<2.5$ is found at $\sqrt{s}=900$~GeV in PYTHIA.

Charged particle transverse momentum distributions
can be used to tune Monte Carlo event generators of hadron-hadron 
interactions, such as PYTHIA and PHOJET. 
Recently, PYTHIA was 
tuned to describe the energy dependence of existing measurements, 
e.g. with respect to the treatment of multiple parton interactions
and divergencies of the 2$\rightarrow$2 parton scattering cross-section 
at small momentum transfers.

\begin{figurehere}
\centering
\includegraphics[width=7.8cm]{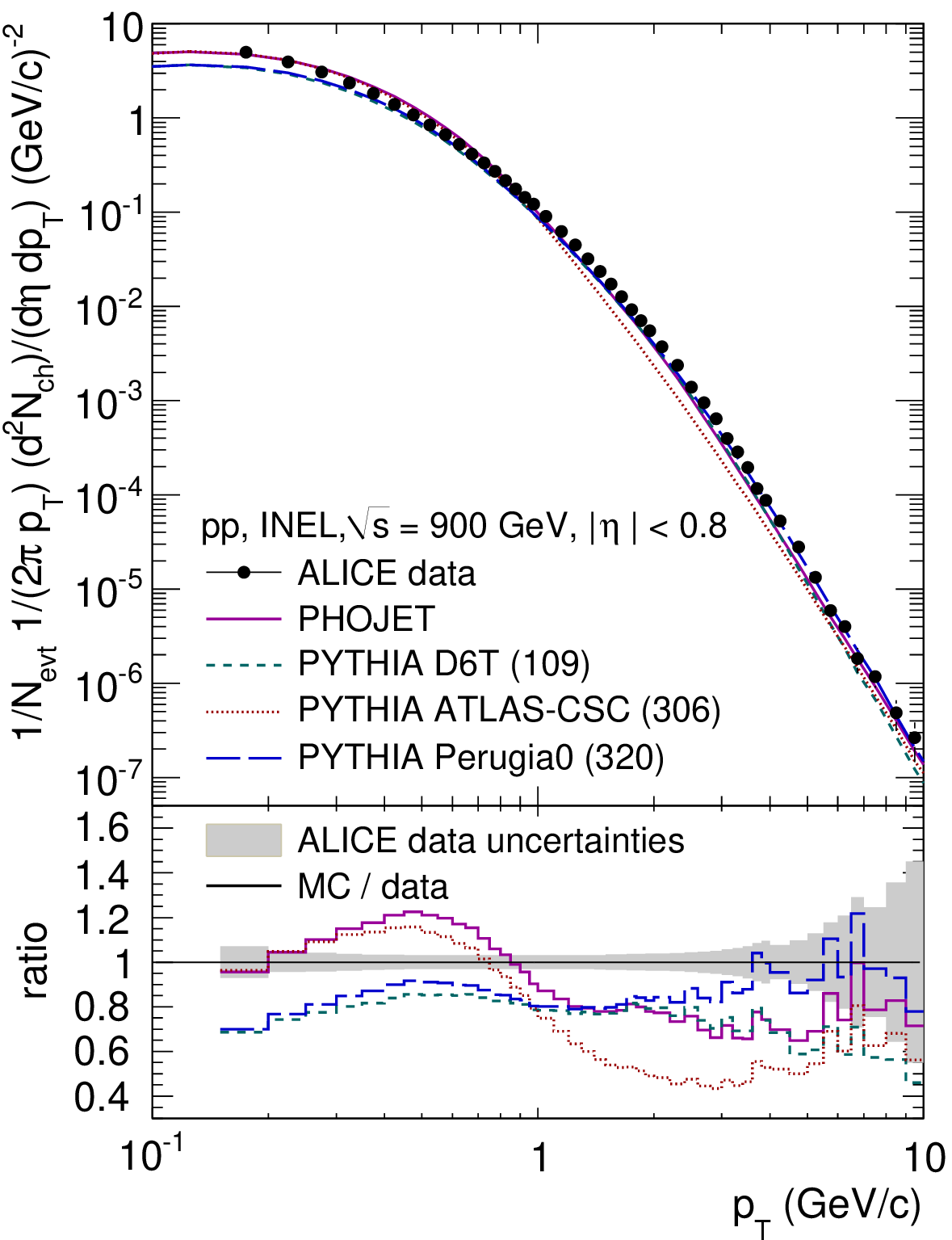}
\caption{\label{inel-models} Top: Comparison of the primary charged particle 
differential yield in INEL $pp$ collisions at $\sqrt{s}=900$~GeV 
($|\eta|<0.8$) to results from PHOJET and PYTHIA tunes 109~\cite{109-tune},
306~\cite{306-tune} and 320~\cite{320-tune}. 
Bottom: Ratio between the Monte Carlo simulation and the data. The shaded area indicates 
the statistical and 
systematic errors of the ALICE data added in quadrature.}
\end{figurehere}

In Fig.~\ref{inel-models}, the results for INEL events are compared to 
PHOJET and different tunes of PYTHIA, D6T 
(tune~109) \cite{109-tune},
Perugia0 (tune~320)~\cite{320-tune} and ATLAS-CSC 
(tune 306)~\cite{306-tune}. 
The best agreement is found with the Perugia0 tune, which gives
a fair description of the spectral shape, but is approximately 20\%
below the data. The D6T tune is similar to Perugia0 below 2~GeV/$c$ but 
underestimates the data more significantly at high $p_T$. 
PHOJET and the PYTHIA ATLAS-CSC tune fail to reproduce the
spectral shape of the data. They overestimate the yield below 0.7~GeV/$c$
and fall short of the data at high $p_T$.
We note that PHOJET and ATLAS-CSC agree best 
with the charged particle multiplicity distributions at $\sqrt{s}~=~$0.9, 2.36
 and 7~TeV, respectively~\cite{paper2,paper7tev}.

Fig.~\ref{spectra-mult} shows the $p_T$ spectra in
INEL events for three different multiplicity selections 
$(n_{\rm acc})$ along with fits to the modified Hagedorn
function (Eq.~\ref{param}).
A considerable flattening of the tails of the spectra is visible
with increasing multiplicity. The fit parameters $p_{T,0}$ and $b$ drop
by more than 50\% from the lowest to the highest multiplicities.
The results for the fit parameters in bins of $n_{\rm acc}$ are listed in 
Table~\ref{tab2}, along with the average multiplicity
$\langle n_{\rm ch}\rangle$ assigned to each $n_{\rm acc}$ as determined
from Monte Carlo simulations.

\begin{figurehere}
\centering
\includegraphics[width=7.8cm]{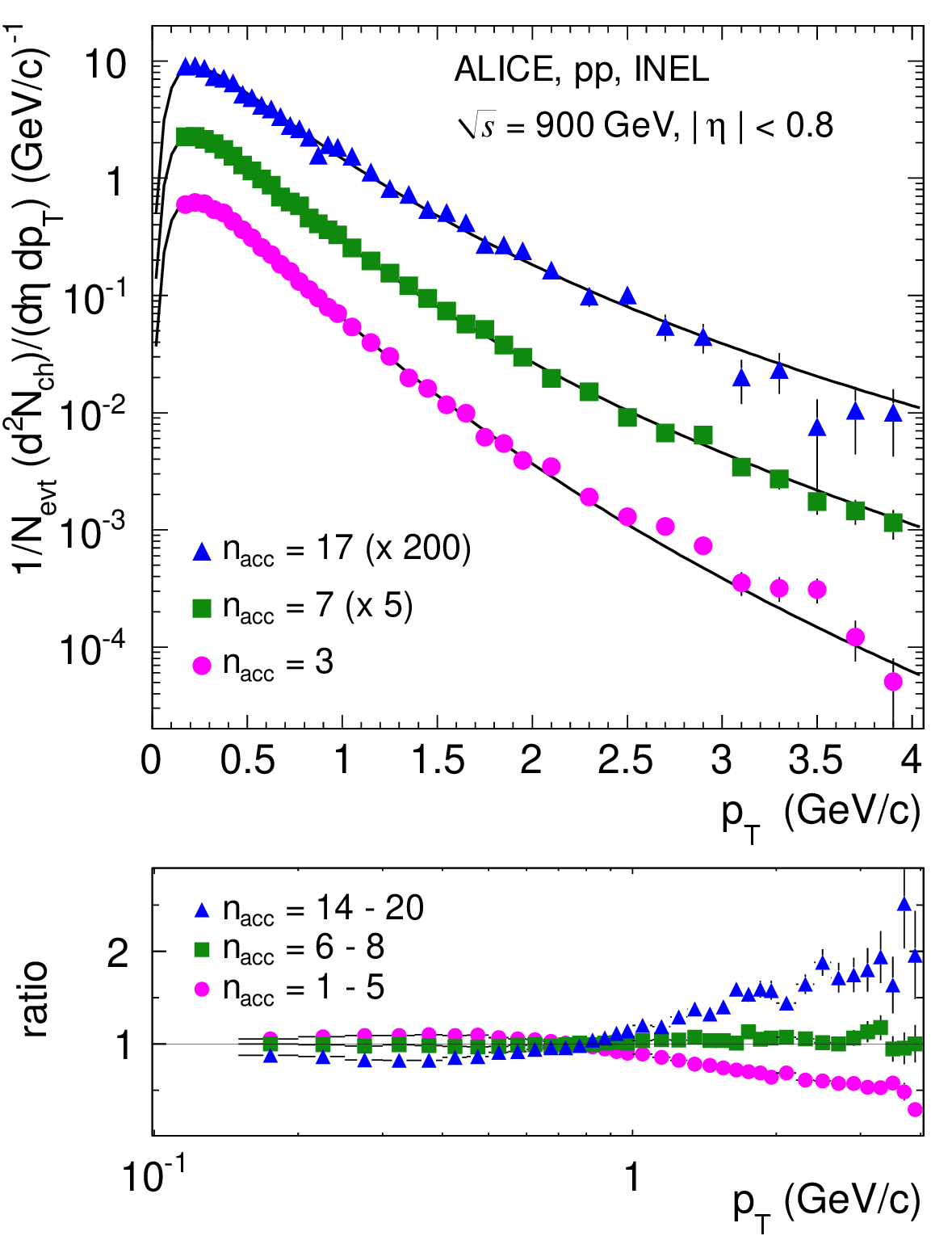}
\caption{\label{spectra-mult} Upper panel: Transverse momentum spectra of primary 
charged particles in INEL $pp$ collisions at $\sqrt{s}=900$~GeV 
($|\eta|<0.8$), normalized to the total number of INEL events $N_{\rm evt}$, for three different 
event multiplicities  together 
with the modified Hagedorn fits (Eq.~\ref{param}) described in the text. 
The fits are performed in the range $0.15 < p_T < 4$~GeV/$c$ and extrapolated
to $p_T=0$. The
error bars indicate the statistical and systematic errors added
in quadrature. Lower panel: Ratios of the $p_T$
spectra in different multiplicity ranges to the inclusive $p_T$
spectrum in INEL events.}
\end{figurehere}

Also shown in Fig.~\ref{spectra-mult} are ratios of $p_T$
spectra in different multiplicity regions over the inclusive $p_T$
spectrum in INEL events. A very pronounced multiplicity dependence of
the spectral shape is manifest, exhibiting enhanced particle production
at high $p_T$ in high multiplicity events. At $p_T<0.8$~GeV/$c$ the trend is 
opposite, albeit with a much weaker multiplicity dependence.
The evolution of the spectral shape with multiplicity may shed light
on different particle production mechanisms in $pp$ collisions.
A qualitatively similar evolution of the $p_T$ spectra with 
multiplicity has been seen in $pp$ data at $\sqrt{s}=200$~GeV~\cite{lisa}.

\begin{figure*}
\hspace{0.5cm}
\subfigure{\includegraphics[width=7.0cm]{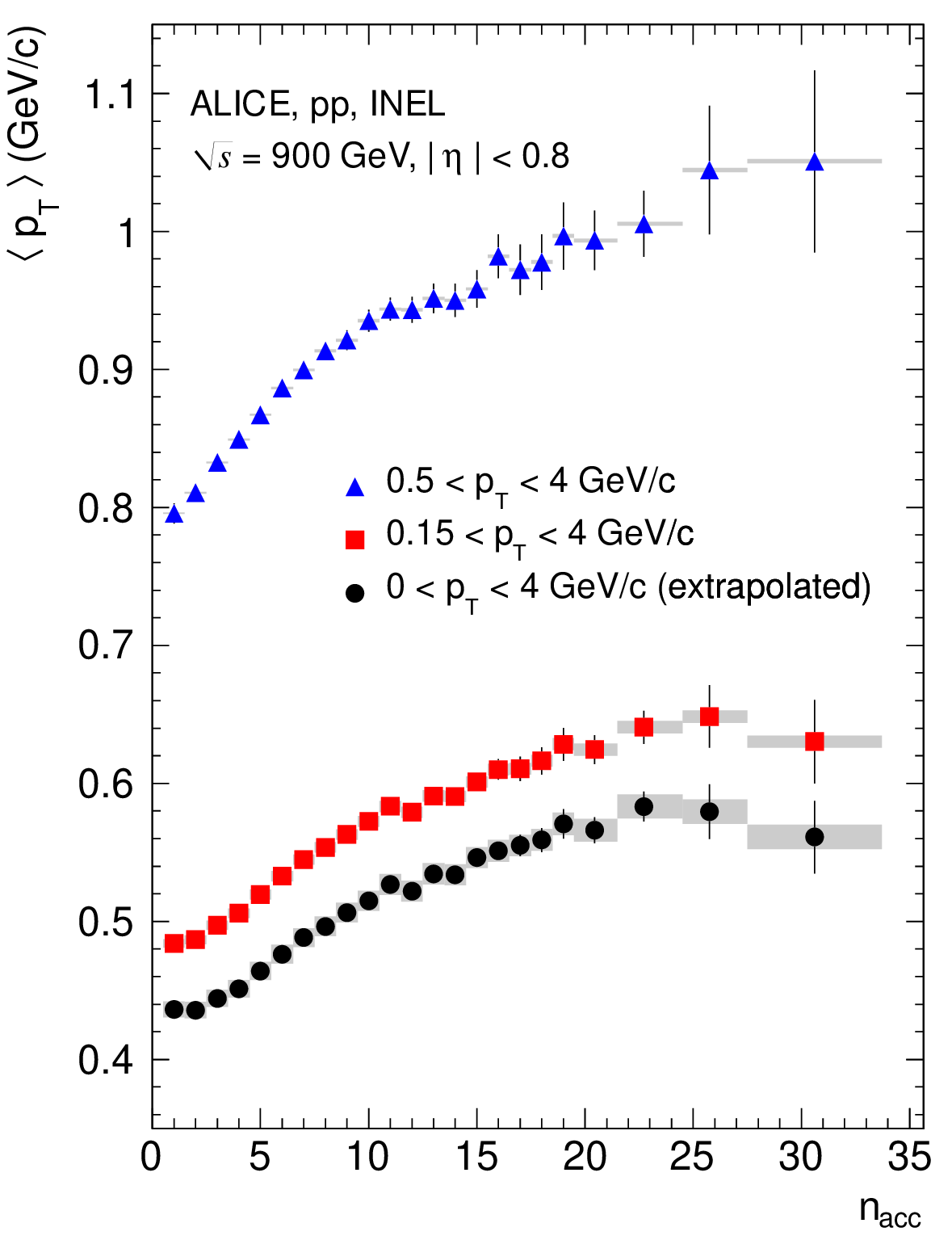}} \hspace{1.0cm}
\subfigure{\includegraphics[width=7.0cm]{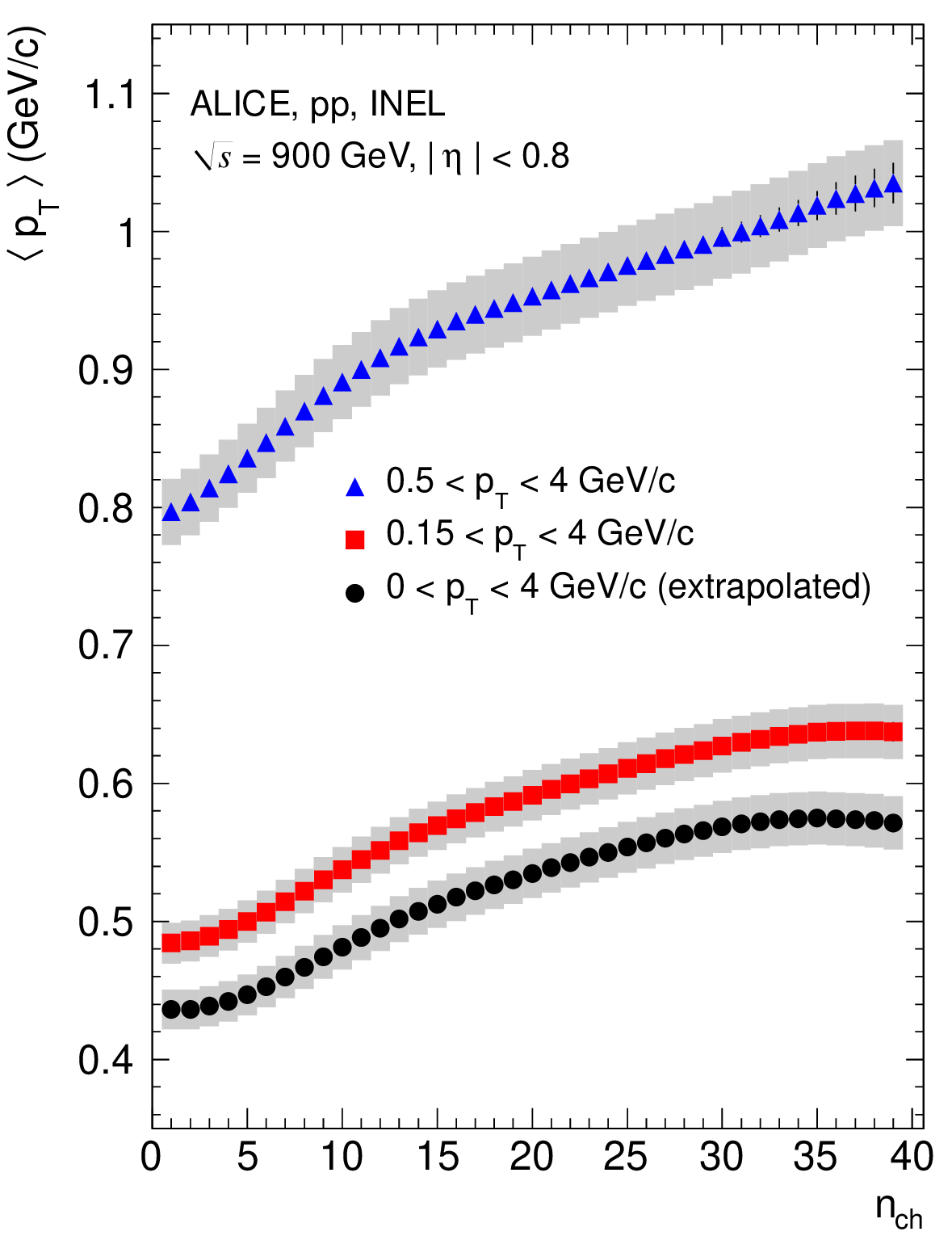}}
\caption{\label{mpt-n} The average transverse momentum of 
charged particles in INEL $pp$ events at $\sqrt{s}=900$~GeV 
for three different $p_T$ ranges
as a function of $n_{\rm acc}$ (left panel) and as a 
function of $n_{\rm ch}$ (right panel). 
The error bars and shaded areas indicate the statistical and 
systematic errors, respectively.} 
\hspace{0.5cm}
\subfigure{\includegraphics[width=7.0cm]{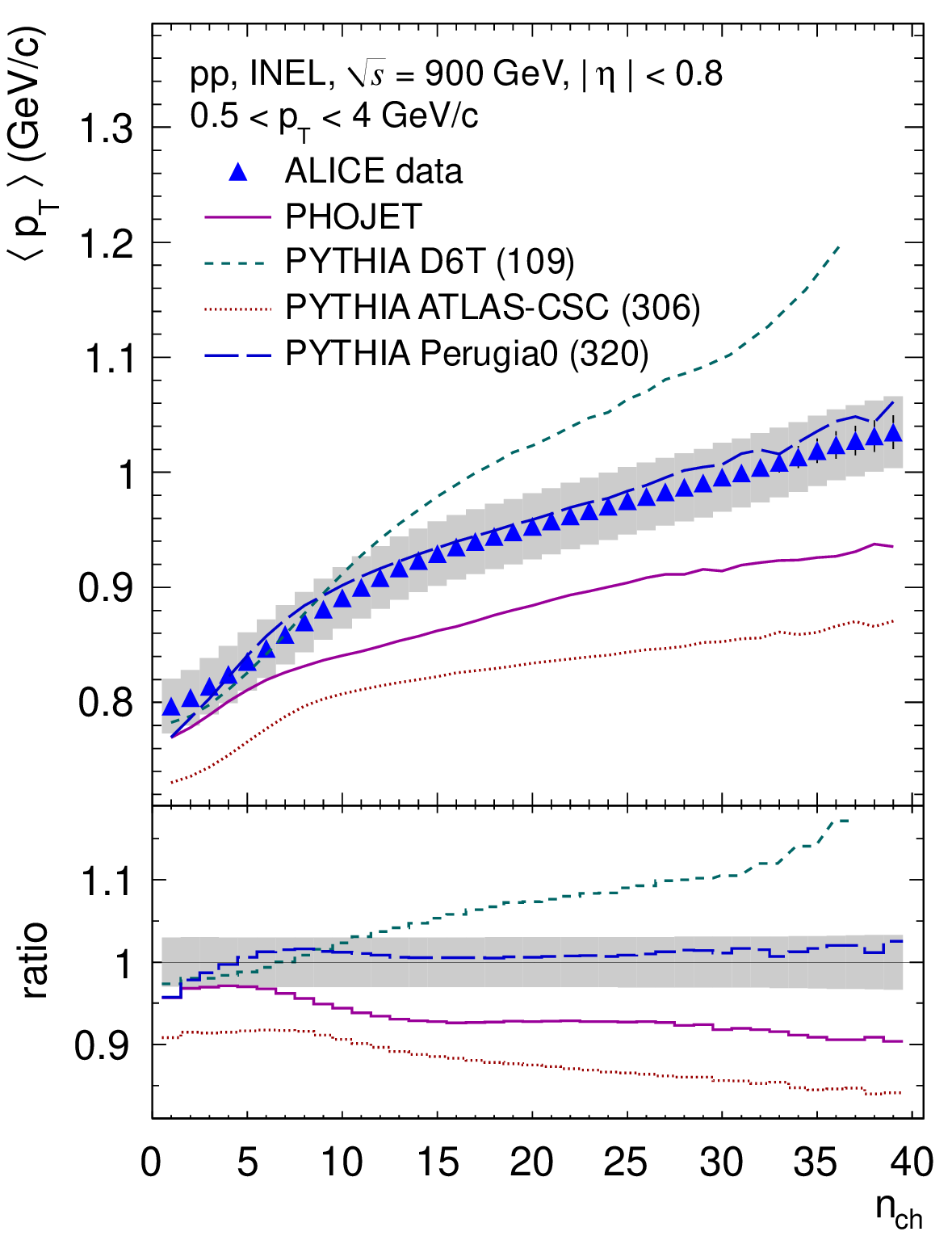}}\hspace{1.0cm}
\subfigure{\includegraphics[width=7.0cm]{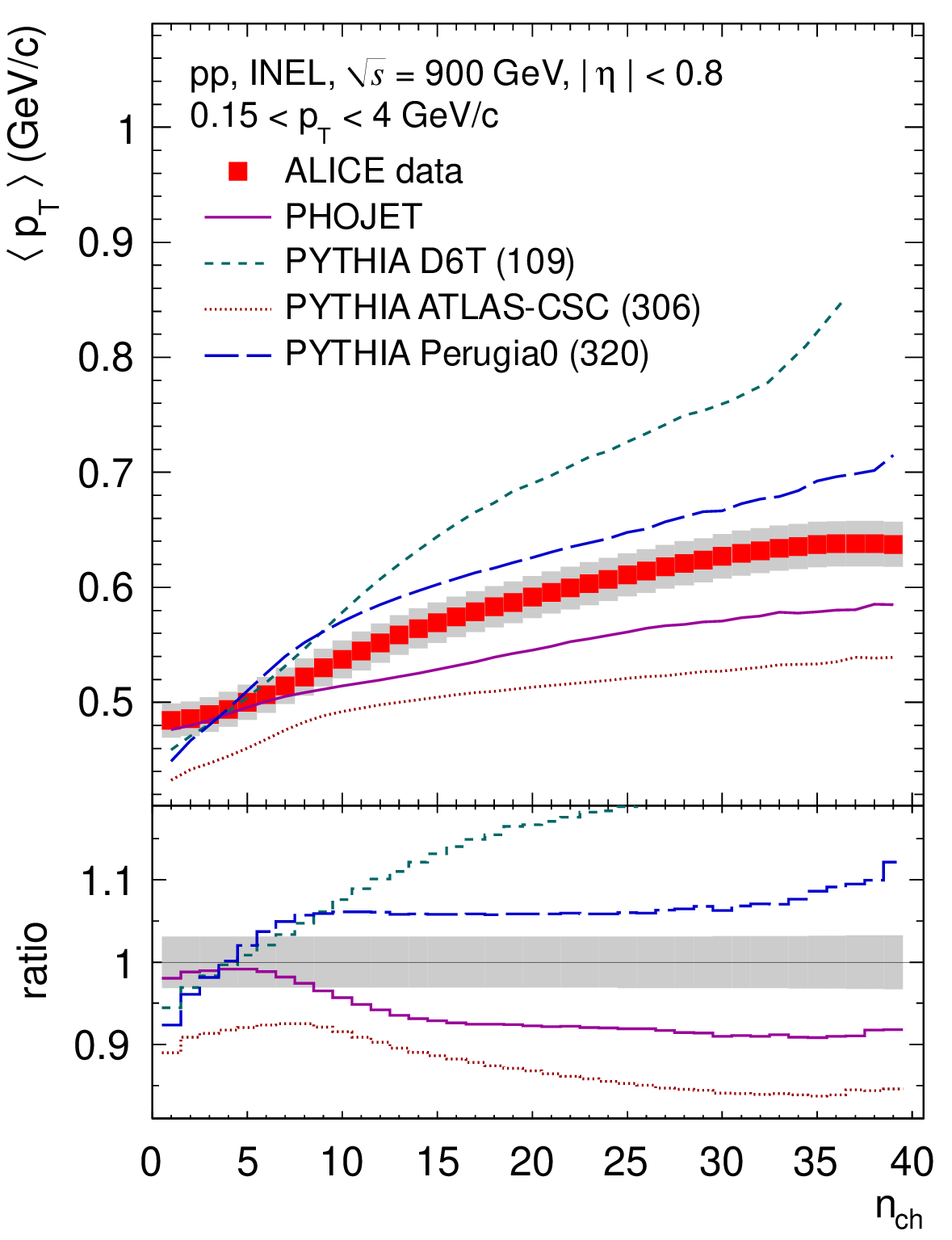}}
\caption{\label{mpt-models}   
The average transverse momentum of charged particles for $0.5<p_T<4$~GeV/$c$ 
(left panel) and $0.15<p_T<4$~GeV/$c$ (right panel) in INEL $pp$ events at 
$\sqrt{s}=900$~GeV as a function 
of $n_{\rm ch}$ in comparison to models. The error bars and the shaded area
indicate the statistical and systematic errors of the data, respectively.
In the lower panels, the ratio Monte Carlo over data is shown. 
The shaded areas 
indicate the statistical and systematic uncertainty of the data,
added in quadrature.}

\end{figure*}

The average transverse momentum \mpt as a function of the
multiplicity of accepted particles $(n_{\rm acc})$ in INEL
$pp$ collisions at $\sqrt{s}=900$~GeV is shown 
in the left panel of Fig.~\ref{mpt-n}.
For all three selected $p_T$ ranges a significant 
increase of \mpt with multiplicity is observed. Most significantly
for  $0.5 < p_T < 4$~GeV/$c$, the slope changes at intermediate
multiplicities.

In the right panel of Fig.~\ref{mpt-n} the same data is shown 
as a function of $n_{\rm ch}$
after application of the weighting procedure (Eq.~\ref{weighting}).
In comparison to model calculations, good agreement with the data
for $0.5<p_T<4$~GeV/$c$ is found only for the PYTHIA Perugia0 tune 
(Fig.~\ref{mpt-models}, left panel).
In a wider pseudorapidity interval ($|\eta|<2.5$), similar agreement
of the data with Perugia0 was reported by ATLAS~\cite{atlas-paper1}.
For $0.15<p_T<4$~GeV/$c$, Perugia0 and PHOJET are the closest to the data, 
as shown in the right panel of Fig.~\ref{mpt-models}, 
however, none of the models
gives a good description of the entire measurements.

\section{Conclusion}
A measurement is presented of the primary charged particle transverse
momentum spectrum and of the mean transverse momentum 
in $pp$ collisions 
at $\sqrt{s}=900$~GeV with the ALICE detector at the LHC. 
Good agreement with previous results from LHC is found up to $p_T= 1$~GeV/$c$.
At higher $p_T$,
the data exhibit a harder momentum
spectrum of primary charged particles
than other measurements in $pp$ and $p\bar{p}$ collisions
at the same energy.
We argue that this is most likely related to the different pseudorapidity 
intervals studied.
The average transverse momentum in $|\eta|<0.8$ is 
\mpt$_{\rm INEL}=0.483\pm0.001$~(stat.)~$\pm0.007$~(syst.)~GeV/$c$ and  
\mpt$_{\rm NSD}=0.489\pm0.001$~(stat.)~$\pm0.007$~(syst.)~GeV/$c$. 
None of the models and tunes investigated simultaneously describes
the $p_T$ spectrum and the correlation between \mpt and $n_{\rm ch}$.
In particular in the low $p_T$ region, where the bulk of the 
particles are produced, the models require further tuning.
These measurements will help to improve
the phenomenological description of soft QCD processes and the
interplay between soft and hard QCD. The presented data 
demonstrate the excellent performance of the ALICE detector for 
momentum measurement
and will be used as a baseline for measurements at higher LHC energies and for
comparison with particle production in heavy-ion collisions.

\begin{table*}
\centering
\caption{\label{tab1} Contributions to the systematic uncertainties on the 
differential primary charged particle yields 
$1/N_{\rm evt}~1/(2\pi p_T)~{\rm d}^2N_{\rm ch}/({\rm d}\eta~{\rm d}p_T)$ 
and the average transverse momentum \mpt.
Ranges are given if the contributions are $p_T$ dependent.}
~\\
~\\
\begin{tabular}{ll|c|ccc}
   ~&~&$\frac{1}{N_{\rm evt}}\frac{1}{2\pi p_T}\frac{{\rm d}^2N_{\rm ch}}{{\rm d}\eta~{\rm d}p_T}$  &~&\mpt&~\\
\hline 
   ~&$p_T$ range (GeV/$c$) & $0.15-10$ &$0.5-4$ &$0.15-4$ &$0-4$~(extrap.) \\
\hline
   ~&Track selection cuts               & 0.2-4\%&~negl.&0.3\%&0.5\%\\
   ~&Contribution of diffraction (INEL) & 0.9-1\%&~negl.&negl.&negl.  \\
   ~&Contribution of diffraction (NSD)  & 2.8-3.9\%&~-&-&-  \\
   ~&Event generator dependence (INEL)  & 2.5\%&~negl.&negl.&negl. \\
   ~&Event generator dependence (NSD)   & 0.5\%&~-&-&- \\
   ~&Particle composition               & 1-2\%&~0.1\%&negl.&0.1\% \\
   ~&Secondary particle rejection	& 0.2-1.5\%&~negl.&0.1\%&0.2\%\\
   ~&Detector misalignement             & negl.&~negl.&negl.&negl. \\
   ~&ITS efficiency			& 0-1.6\% &~negl.&0.3\%&0.5\%\\
   ~&TPC efficiency			& 0.8-4.5\%&~negl.&0.5\%&0.7\%\\
   ~&SPD triggering efficiency		& negl.&~negl.&negl.&negl. \\
   ~&VZERO triggering efficiency (INEL) & negl.&~negl.&negl.&negl. \\
   ~&VZERO triggering efficiency (NSD)  & 0.2\%&~-&-&- \\
   ~&Beam-gas events                    & negl.&~negl.&negl.&negl. \\
   ~&Pile-up events                     & negl.&~negl.&negl.&negl. \\
\hline
   ~&Total (INEL)                         & 3.0-7.1\%&~0.1\%&0.7\%&1.0\% \\
   ~&Total (NSD)                          & 3.5-7.2\%&~-&-&- \\
\hline
  ~&$R$ weighting procedure		& ~& 3.0\% & 3.0\% & 3.0\%\\
  ~&Extrapolation to $p_T=0$		& ~& - & - & 1.0\% \\
\hline
  ~&Total 				& ~&3.0\%& 3.1\%&3.3\%

\end{tabular}
\end{table*} 

\begin{table*}
\centering
\caption{\label{tab2} Parameters of the modified Hagedorn fits (Eq.~\ref{param}) to 
the transverse momentum spectra. The fit range in $p_T$ is $0.15-10$~GeV/$c$ 
for the multiplicity integrated spectra (first two rows) and 
$0.15-4$~GeV/$c$ for the spectra binned in multiplicity. The errors are 
statistical and systematic added in
quadrature. Also given are the average multiplicites $\langle n_{\rm ch}\rangle$ of events
contributing to the $n_{\rm acc}$ bins, as determined from Monte Carlo.}
~\\
~\\
\begin{tabular}{ccccc}
event class 	 	& $n_{\rm acc}$ & $\langle n_{\rm ch}\rangle$ & $p_{T,0}$ (GeV/$c$)	& $b$ \\
\hline
INEL 	& all 	&~	&$1.05\pm0.05$		&  $7.92\pm0.04$ \\
NSD	& all	&~	&$1.05\pm0.05$		&  $7.84\pm0.04$ \\
\hline
INEL	&	1 & $2.1\pm0.1$ &$2.64\pm0.29$		&  $16.50\pm1.32$ \\
INEL	&	2 & $3.5\pm0.1$ &$1.86\pm0.15$		&  $12.58\pm0.69$ \\
INEL	&	3 & $4.8\pm0.1$ &$1.49\pm0.11$		&  $10.56\pm0.45$ \\
INEL	&	4 & $6.1\pm0.1$ &$1.26\pm0.08$		&  $9.28\pm0.34$ \\
INEL	&	5 & $7.4\pm0.1$ &$1.16\pm0.07$		&  $8.60\pm0.28$ \\
INEL	&	6 & $8.7\pm0.1$ &$1.04\pm0.06$		&  $7.87\pm0.24$ \\
INEL	&	7 & $10.0\pm0.2$ &$1.01\pm0.07$		&  $7.60\pm0.23$ \\
INEL	&	8 & $11.3\pm0.2$ &$0.95\pm0.05$		&  $7.27\pm0.21$ \\
INEL	&	9 & $12.6\pm0.2$ &$0.97\pm0.06$		&  $7.28\pm0.22$ \\
INEL	&	10 & $13.9\pm0.3$ &$0.90\pm0.06$		&  $6.87\pm0.21$ \\
INEL	&	11 & $15.1\pm0.3$ &$0.91\pm0.06$		&  $6.82\pm0.21$ \\
INEL	&	12 & $16.4\pm0.3$ &$0.90\pm0.06$		&  $6.80\pm0.22$ \\
INEL	&	13 & $17.7\pm0.4$ &$0.91\pm0.06$		&  $6.74\pm0.23$ \\
INEL	&	14 & $18.9\pm0.5$ &$0.89\pm0.06$		&  $6.65\pm0.24$ \\
INEL	&	15 & $20.1\pm0.5$ &$0.96\pm0.07$		&  $6.88\pm0.27$ \\
INEL	&	16 & $21.3\pm0.6$ &$0.79\pm0.06$		&  $6.14\pm0.23$ \\
INEL	&	17 & $22.5\pm0.5$ &$0.92\pm0.08$		&  $6.64\pm0.30$ \\
INEL	&	18 & $23.7\pm0.6$ &$0.84\pm0.08$		&  $6.29\pm0.29$ \\
INEL	&	19 & $24.9\pm0.7$ &$0.80\pm0.09$		&  $6.06\pm0.31$ \\
INEL	&	20-21 & $26.6\pm0.7$ &$0.79\pm0.09$		&  $6.03\pm0.31$ \\
INEL	&	22-24 & $29.4\pm0.8$ &$0.78\pm0.09$		&  $5.89\pm0.32$ \\
INEL	&	25-27 & $33.0\pm1.1$ &$0.54\pm0.10$		&  $5.02\pm0.40$ \\
INEL	&	28-45 & $37.1\pm1.5$ &$0.59\pm0.16$		&  $5.42\pm0.67$ \\

\end{tabular}
\end{table*}

\end{multicols}

\section*{Acknowledgements}
The ALICE collaboration would like to thank all its engineers and technicians 
for their invaluable contributions to the construction of the experiment and 
the CERN accelerator teams for the outstanding performance of the LHC complex.

The ALICE collaboration acknowledges the following funding agencies for their support in building and
running the ALICE detector:
\begin{itemize}
\item{}
Calouste Gulbenkian Foundation from Lisbon and Swiss Fonds Kidagan, Armenia;
\item{}
Conselho Nacional de Desenvolvimento Cient\'{\i}fico e Tecnol\'{o}gico (CNPq), Financiadora de Estudos e Projetos (FINEP),
Funda\c{c}\~{a}o de Amparo \`{a} Pesquisa do Estado de S\~{a}o Paulo (FAPESP);
\item{}
National Natural Science Foundation of China (NSFC), the Chinese Ministry of Education (CMOE)
and the Ministry of Science and Technology of China (MSTC);
\item{}
Ministry of Education and Youth of the Czech Republic;
\item{}
Danish Natural Science Research Council, the Carlsberg Foundation and the Danish National Research Foundation;
\item{}
The European Research Council under the European Community's Seventh Framework Programme;
\item{}
Helsinki Institute of Physics and the Academy of Finland;
\item{}
French CNRS-IN2P3, the `Region Pays de Loire', `Region Alsace', `Region Auvergne' and CEA, France;
\item{}
German BMBF and the Helmholtz Association;
\item{}
Hungarian OTKA and National Office for Research and Technology (NKTH);
\item{}
Departments of Atomic Energy and Science and Technology, Government of India;
\item{}
Istituto Nazionale di Fisica Nucleare (INFN) of Italy;
\item{}
MEXT Grant-in-Aid for Specially Promoted Research, Ja\-pan;
\item{}
Joint Institute for Nuclear Research, Dubna;
\item{}
Korea Foundation for International Cooperation of Science and Technology (KICOS);
\item{}
CONACYT, DGAPA, M\'{e}xico, ALFA-EC and the HELEN Program (High-Energy physics Latin-American--European Network);
\item{}
Stichting voor Fundamenteel Onderzoek der Materie (FOM) and the Nederlandse Organisatie voor Wetenschappelijk Onderzoek (NWO), Netherlands;
\item{}
Research Council of Norway (NFR);
\item{}
Polish Ministry of Science and Higher Education;
\item{}
National Authority for Scientific Research - NASR (Autontatea Nationala pentru Cercetare Stiintifica - ANCS);
\item{}
Federal Agency of Science of the Ministry of Education and Science of Russian Federation, International Science and
Technology Center, Russian Academy of Sciences, Russian Federal Agency of Atomic Energy, Russian Federal Agency for Science and Innovations and CERN-INTAS;
\item{}
Ministry of Education of Slovakia;
\item{}
CIEMAT, EELA, Ministerio de Educaci\'{o}n y Ciencia of Spain, Xunta de Galicia (Conseller\'{\i}a de Educaci\'{o}n),
CEA\-DEN, Cubaenerg\'{\i}a, Cuba, and IAEA (International Atomic Energy Agency);
\item{}
Swedish Reseach Council (VR) and Knut $\&$ Alice Wallenberg Foundation (KAW);
\item{}
Ukraine Ministry of Education and Science;
\item{}
United Kingdom Science and Technology Facilities Council (STFC);
\item{}
The United States Department of Energy, the United States National
Science Foundation, the State of Texas, and the State of Ohio.
\end{itemize}

\end{document}